\documentclass[10pt]{article} \usepackage{aaspp4}

\newcommand{\CII}{C$\;${\small\rm II}\relax}
\newcommand{\NI}{N$\;${\small\rm I}\relax}
\newcommand{\OI}{O$\;${\small\rm I}\relax}

\newcommand{\SiII}{Si$\;${\small\rm II}\relax}
\newcommand{\SII}{S$\;${\small\rm II}\relax}

\newcommand{\FeII}{Fe$\;${\small\rm II}\relax}

\newcommand{\lya}{Ly$\alpha$}

\newcommand{\wave}[1]{$\lambda$#1\relax}

\newcommand{\kms}{km~s$^{-1}$\relax}
\newcommand{\etal}{{\em et al.}\relax}

\newcommand{\column}{cm$^{-2}$}

\newcommand{\eij}{$e_{i,j}$}
\newcommand{\gij}{$g_{i,j}$}
\newcommand{\bij}{$b_{i,j}$}

\newcommand{\hd}{HD~}

\newcommand{\hst}{{\em HST}}
\newcommand{\ghrs}{GHRS}
\newcommand{\stis}{STIS}
\newcommand{\calstis}{{\tt CALSTIS}}

\begin{document}


\title{Background and Scattered Light Subtraction in the High-Resolution 
Echelle Modes of the Space Telescope Imaging Spectrograph\altaffilmark{1}}

\altaffiltext{1}{Based on observations made with the NASA/ESA Hubble Space
Telescope, obtained from the data archive at the Space Telescope
Science Institute. STScI is operated by the Association of Universities
for Research in Astronomy, Inc. under the NASA contract NAS 5-26555. }

\author{J. Christopher Howk \& Kenneth R. Sembach}

\affil{Department of Physics and Astronomy\\
The Johns Hopkins University\\ Baltimore, MD, 21218\\
howk@pha.jhu.edu, sembach@pha.jhu.edu}

\authoremail{howk@pha.jhu.edu}

\begin{center}
To appear in {\em The Astronomical Journal} \\ 
\end{center}
\vspace{0.15in}


\begin{abstract}

We present a simple, effective approach for estimating the on-order
backgrounds of spectra taken with the highest-resolution modes of the
Space Telescope Imaging Spectrograph (\stis) on-board the {\em Hubble
Space Telescope}.  Our scheme for determining the on-order background
spectrum for STIS E140H and E230H observations uses moderate-order
polynomial fits to the inter-order scattered light visible in the
two-dimensional STIS MAMA images.  We present a suite of
high-resolution STIS spectra to demonstrate that our background
subtraction routine produces the correct overall zero point, as judged
by the small residual flux levels in the centers of strongly-saturated
interstellar absorption lines.  Although there are multiple sources of
background light in STIS echelle mode data, this simple approach works
very well for wavelengths longward of Lyman-$\alpha$ ($\lambda \ga
1215$ \AA).  At shorter wavelengths, the smaller order separation and
generally lower signal-to-noise ratios of the data can reduce the
effectiveness of our background estimation procedure.  Slight
artifacts in the background-subtracted spectrum can be seen in some
cases, particularly at wavelengths $\la 1300$ \AA.  Most of these are
caused by echelle scattering of strong spectral features into the
inter-order light.  We discuss the limitations of high-resolution STIS
data in light of the uncertainties associated with our background
subtraction procedure.  We compare our background-subtracted STIS
spectra with GHRS Ech-A observations of the DA white dwarf G191-B2B
and GHRS first-order G160M observations of the early-type star
HD~218915.  We find no significant differences between the GHRS data
and the STIS data reduced with our method in either case.
\end{abstract}



\section{Introduction}

The Space Telescope Imaging Spectrograph (\stis) is a
second-generation instrument on the {\em Hubble Space Telescope}
(\hst) designed to offer a wide range of spectroscopic capabilities
from ultraviolet ($\lambda \ga 1150$ \AA) to near-infrared ($\lambda
\la 11,000$ \AA) wavelengths.  The design and construction of \stis\
are described by Woodgate \etal\ (1998), while information about the
on-orbit performance of \stis\ is summarized by Kimble \etal\ (1998).

Spectra taken in the visible wavelength regime ($\lambda \ga 3000$
\AA) are recorded by a large-format (1024$\times$1024 pixels) CCD
detector, while the ultraviolet (UV) portion of the spectrum is
covered by two photon-counting multianode microchannel array (MAMA)
detectors.  The anode arrays used in the MAMA detectors contain
1024$\times$1024 elements, but photon events are centroided to half
the natural spacing of the anode arrays. The detectors are read out as
2048$\times$2048 pixel arrays, but for our purposes (and in most
\stis\ documentation, see Kimble \etal\ 1998) a pixel will be defined
as one element in the lower-resolution 1024$\times$1024 element format.
Each of the two MAMA detectors are used (in their primary modes) to
image a separate portion of the UV spectrum.  The far-ultraviolet
(FUV) MAMA covers the wavelength range $1150-1700$ \AA, while the
near-ultraviolet (NUV) MAMA is primarily used in the wavelength range
$1650-3100$ \AA.  The two-dimensional format of the \stis\ detectors
offers significant multiplexing advantages over the one-dimensional
Digicon detectors used in the two previous-generation spectrographs on
\hst, the Goddard High Resolution Spectrograph (GHRS) and the Faint
Object Spectrograph.

High-resolution UV spectroscopic capabilities with \stis\ are provided
by four cross-dispersed echelle gratings, yielding resolutions
$R=\lambda/\Delta \lambda$ in the range 30,000--110,000 with the
primary apertures (the $0\farcs2\times0\farcs09$ or
$0\farcs2\times0\farcs2$ apertures for the E140H and E230H gratings,
and $0\farcs2\times0\farcs06$ or $0\farcs2\times0\farcs2$ apertures
for the E140M and E230M gratings).  Higher resolution capabilities (up
to $R\sim220,000$) are available using the smallest available entrance
aperture ($0\farcs1\times0\farcs03$) with the E140H and E230H gratings.

The two-dimensional format of the MAMA detectors coupled with the
$\sim2.75$ \kms\ velocity resolution of the highest-resolution echelle
gratings (E140H and E230H) provide spectroscopic capabilities that are
in many ways superior to those of the \ghrs.  The Ech-A and Ech-B
gratings used with the $1\times512$ pixel Digicon detectors of the
\ghrs\ covered $7-15$ \AA\ per exposure at a resolution $\sim3.3-3.5$
\kms, while \stis\ provides  $\ga200$ \AA\ coverage per
exposure at slightly higher resolution.  Furthermore, the
two-dimensional \stis\ detectors image the on-order spectrum and
inter-order background and scattered light simultaneously, whereas
observations with the \ghrs\ typically required additional overhead to
measure the inter-order background.

The background and scattered light properties of the \ghrs\ were well
studied in both the pre- and post-flight epochs by Cardelli, Ebetts,
\& Savage (1990, 1993).  An appropriate, simple background subtraction 
technique was developed by these authors and subsequently applied in
the standard \ghrs\ data processing pipeline.  The scattered light
properties of \stis\ have been discussed by Landsman \& Bowers (1997)
and Bowers \etal\ (1998).  The \stis\ Instrument Definition Team (IDT)
has developed a complex algorithm for deriving the on-order background
spectrum for the \stis\ echelle-mode data (Bowers \etal\ 1998).  This
algorithm, to be presented by Bowers \& Lindler (in prep.), derives
the scattered light estimate over the whole MAMA field using an
initial estimate of the on-order spectrum, laboratory measurements of
the scattering properties of the gratings, and models of several
sources of background light (including the detector and image point
spread function halos).  The final on-order background is estimated
after iteratively converging on a model image that best matches the
observed MAMA image, and then subtracting the background portion of
the model from the original image.  The implementation of this
background-correction routine into the standard
\calstis\ processing pipeline\footnote{CALSTIS is part of the standard
STSDAS distribution available from Space Telescope Science Institute
(STScI).  More information on CALSTIS and its procedures can be found
in the {\em HST Data Handbook} (Voit 1997).} is currently proceeding.

In this paper we present an alternative approach for deriving the
on-order background spectrum for high-resolution \stis\ echelle data
taken with the E140H and E230H gratings.  This approach is much
simpler and less computationally intensive than that derived by the
\stis\ IDT, and can be used with the data products produced by the
standard \calstis\ distribution.  Our derivation of the on-order
background spectrum is highly empirical; we do not employ detailed
models of the many sources of scattered and background light in the
\stis\ instrument.  This simplicity has its limitations.  The effects of 
scattering by the echelle grating are not explicitly accounted for,
and this causes artifacts in our background at short wavelengths.
However, we will show that our algorithm is able to correctly estimate
the level of scattered light in \stis\ echelle observations, as judged
by the very small residual flux levels in the saturated cores of
strong interstellar absorption lines.  Furthermore, we will present
comparisons of background-corrected \stis\ data with archival \ghrs\
data that suggest the shape of the derived background spectrum is
being estimated correctly.  Artifacts caused by the influence of
echelle-scattered light in our approach are well understood, and the
importance of these difficulties can be sufficiently accounted for in
the final error budget.

We begin by briefly discussing the \stis\ echelle-mode spectral format
and sources of background and scattered light in \stis\ echelle-mode
data in \S \ref{sec:stis}.  We detail our algorithm for estimating the
on-order background in \stis\ E140H and E230H data in \S
\ref{sec:algorithm}. In \S \ref{sec:examples} we provide examples of
background-corrected data and discuss the uncertainties of our
approach.  Comparisons of \stis\ data extracted using our algorithm
with archival high- and intermediate-resolution \ghrs\ data are given
in \S \ref{sec:ghrs}.  We summarize our results in \S
\ref{sec:summary}.  Our focus throughout this paper will be on the
effects of the background subtraction on the analysis of interstellar
absorption lines, since that is primarily why we have developed our
routines.  However, the approach outlined in this paper should be
applicable to most other uses of high-resolution \stis\ data.  The
routines described in this work will be made available to the general
astronomical community.

\section{The STIS Echelle Format and Scattering Geometry}
\label{sec:stis}

\begin{figure}
\plotone{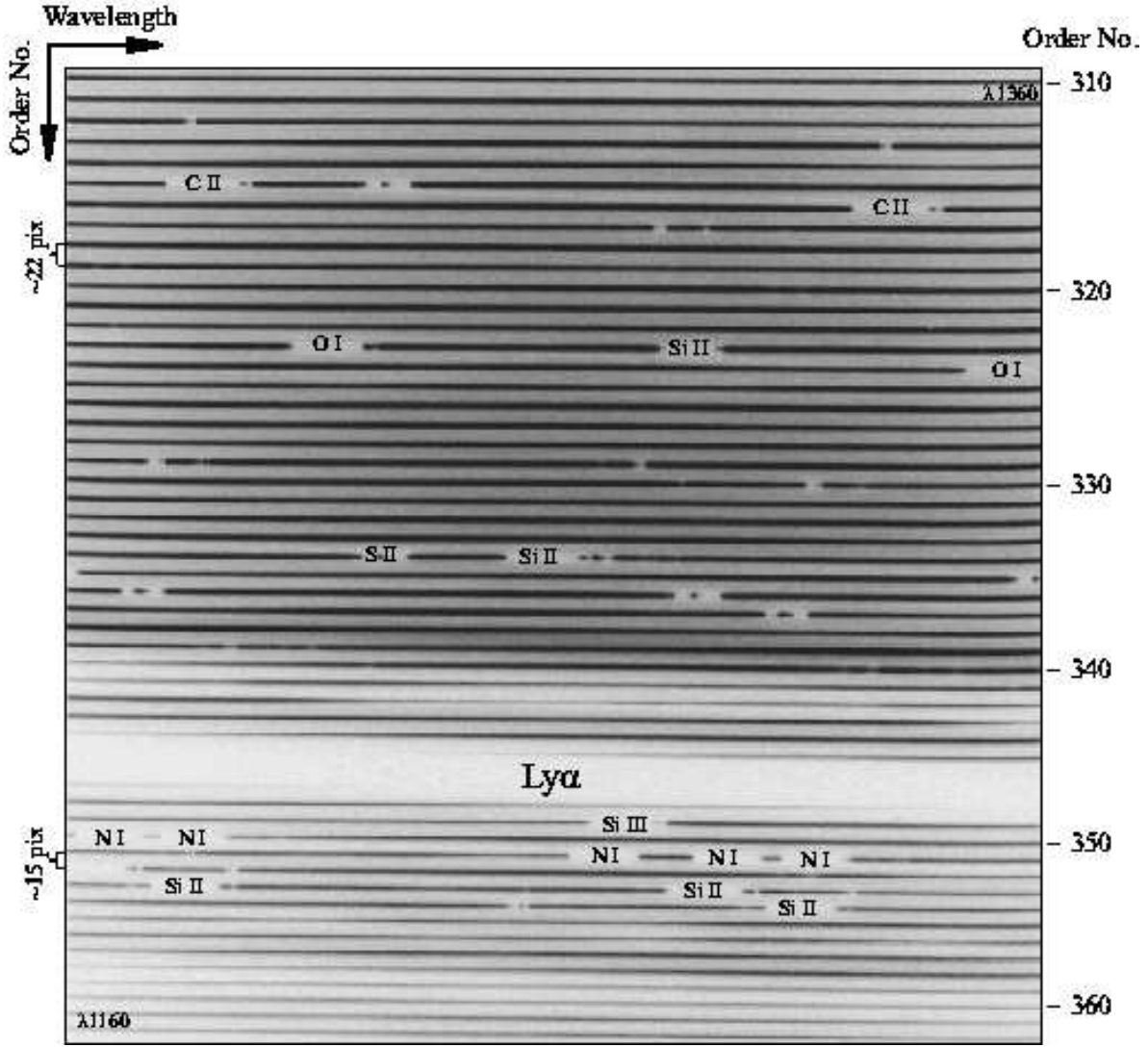}
\caption{A view of a raw MAMA image taken with the STIS E140H grating
and the $0\farcs2\times0\farcs09$ aperture.  Every tenth order is
labelled, and we have identified several strong interstellar
absorption lines.  The target star is HD~218915, which lies behind gas
associated with the Perseus spiral arm.  In this figure, wavelength
increases towards the right and towards the top of the page.  The
total wavelength coverage of this observation is approximately
$1160-1360$ \AA.
\label{fig:fullmama}}
\end{figure}

Figure \ref{fig:fullmama} shows a raw two-dimensional \stis\ E140H FUV
observation of the O9.5 Iab star \hd 218915 covering the wavelength
range $1160-1360$ \AA.  Several prominent interstellar absorption
lines are labelled, and we have marked a number of spectral orders.
The spectral orders at short wavelengths (bottom) are more tightly
spaced than those at the long wavelengths (top).  The usable spectral
orders in this exposure run from 310 to 362 (with central wavelengths
from $\sim1358$ to 1162 \AA, respectively).  Some spectral regions
occur in multiple orders.  For example, two of the three lines in the
\ion{N}{1} triplet near 1200 \AA\ appear in order 350 while the whole
triplet is seen in order 351.  Absorption due to \CII\
\wave{1334} can also be seen in two spectral orders (315 and 316) in
this image.

Figure \ref{fig:fullmama} shows that the individual spectral orders
are tilted with respect to the reference frame defined by the MAMA
detector, and close inspection reveals they are slightly curved.  In
this paper we will be working with a product of the standard \calstis\
processing pipeline: two-dimensional geometrically-rectified images of
each order.  These flux- and wavelength-calibrated images have been
extracted such that they are linear in both the wavelength and spatial
(cross-dispersion) directions.  The rectified images contain both the
on-order spectral region as well as the adjacent inter-order light for
each order.  

\begin{figure}
\epsscale{0.9}
\begin{center}
\plotone{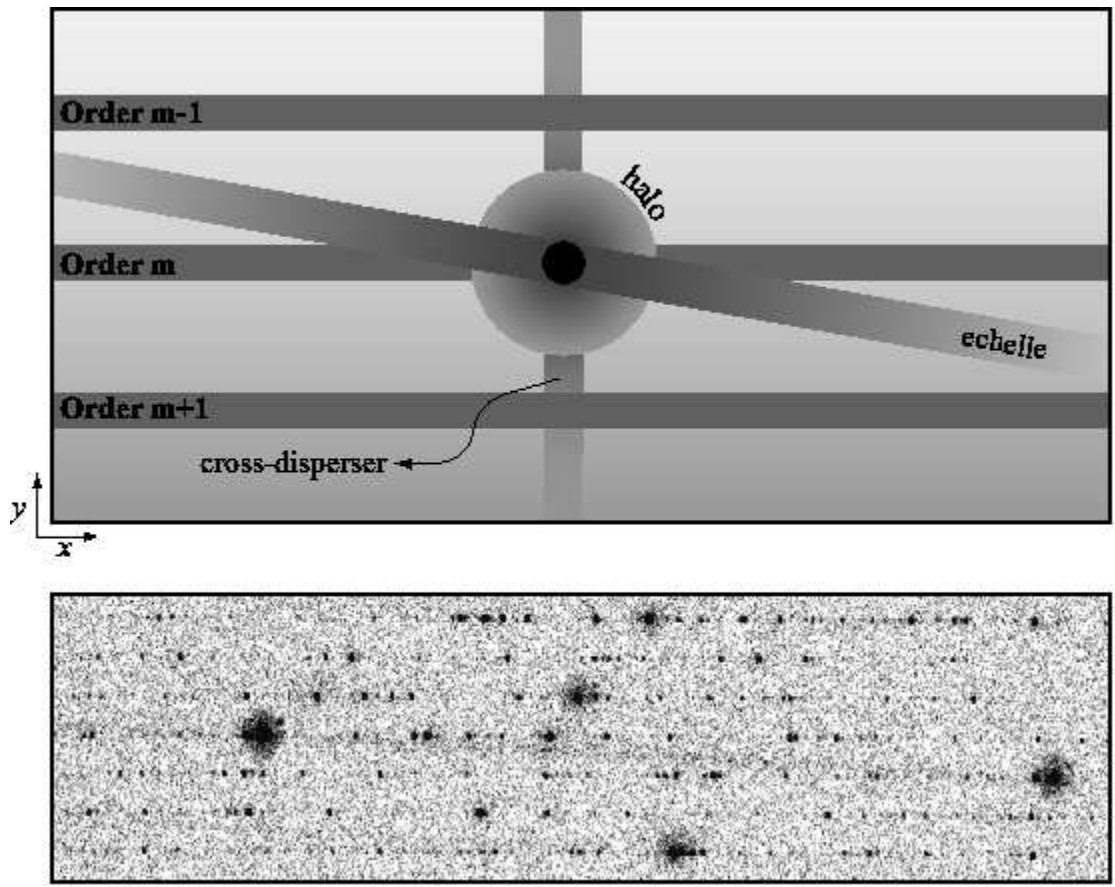}
\end{center}
\caption{{\em Top:} A schematic showing the various possible scattering 
geometries for an emission line within the STIS two-dimensional
echelle data.  The rectified two-dimensional images discussed in the
text include only one order and the inter-order light on either side
of the order, although we have shown the adjacent orders in this
figure.  Of the three primary forms of background light illustrated,
the most important are the effects of the detector halo and echelle
scattering (see text).  There is little evidence for a significant
contribution from cross-disperser scattering (Landsman \& Bowers
1997).  The smaller-scale PSF halo is not illustrated in this figure.
We only show the scattered light from the strong emission line in this
figure; the true background will contain a contribution from the
continuum as well. {\em Bottom:} A section of a deep post-flight STIS
E230H wavelength calibration lamp image.  This region is centered
between 2400 and 2500 \AA, and includes several orders.  Wavelength
increases to the right and towards the top of this image.  (This
section has been flipped across the horizontal axis to be consistent
with the schematic and with Figure 1; E230H images typically have
wavelengths increasing towards the bottom of the image.)  A strong
emission line occurs in both the third and fourth orders from the
bottom of this image section.  The strong halos can be seen
surrounding both images of this particular line, while the faint
signature of echelle scattering can also be discerned.
\label{fig:x2dcartoon}}
\end{figure}

Figure \ref{fig:x2dcartoon} schematically illustrates the geometry of
the two-dimensional rectified images (top panel).  A portion of a
\stis\ comparison lamp observation (taken with the E230H grating) is 
also shown (bottom panel).  We define the $x$ and $y$ axes to follow
the dispersion and cross-dispersion directions, respectively, in these
images.  The center of a spectral trace is defined to be at the
position $y_0$, and the on-order light can be approximately described
by a narrow Gaussian centered on this point.  Though the rectified
images we will be using contain only one order, we have shown several
in Figure \ref{fig:x2dcartoon} for the sake of illustration.

Figure \ref{fig:x2dcartoon} also illustrates several potential forms
of background light in the \stis\ echelle modes, and many of these can
be identified in the accompanying comparison lamp image.  We will
discuss each of these sources of background light below.  Sources of
scattered and background light in the \stis\ echelle modes are also
discussed briefly by Bowers \etal\ (1998) and Landsman \& Bowers
(1997).  For comparison we point the reader to Cardelli \etal\ (1990,
1993) for discussions of the scattering geometry in the \ghrs\ echelle
modes, and to Schroeder (1987) for echelle geometry in general.

Both the echelle and cross-disperser gratings can give rise to
scattered light within the spectrograph.  Cross-disperser scattering
will cause light from the order to be projected perpendicular to the
spectral order, i.e., along the $y$ axis, into the inter-order region.
Cross-disperser scattering is thought to be a relatively minor
contributor in the \stis\ echelle modes (Bowers
\etal\ 1998; Landsman \& Bowers 1997).

Echelle scattering will cause light to be dispersed into the
inter-order region at small angles to the order along lines of constant
wavelength.  For example, a bright emission line will appear to be
connected to the same wavelength in adjacent orders by echelle
scattered light in the inter-order region (see Figure
\ref{fig:x2dcartoon}).  This form of scattering can be an important 
contributor to the background light in the \stis\ high-resolution
echelle modes.  It is difficult to account for without an {\em a
priori} knowledge of the incoming spectrum.  The iterative approach
adopted by the \stis\ IDT should be able to account for this form of
scattered light more accurately than our approach outlined in \S
\ref{sec:algorithm}.

Another important source of background light in the \stis\ echelle
mode data is the detector ``halo,'' whose width for the MAMA detectors
can be of the same magnitude as the separation between orders
(Landsman \& Bowers 1997).  The background spectrum caused by the
detector halo will include contributions from the on-order spectrum of
interest, $m$, and potentially light from the adjacent orders,
$m\pm1$, as well.  Unlike light scattered by the cross-disperser and
echelle gratings, the detector halo creates a background with an
(approximately) axisymmetric pattern.

Similarly, the halo of the telescope point spread function (PSF) will
be an additional source of background light. (This form of background
light is not illustrated in Figure \ref{fig:x2dcartoon}).  This is
essentially light from the wings of the PSF that arrives at the
gratings with a slightly different angle than the narrow core of the
PSF.  Thus the halo creates its own, lower resolution version of the
primary on-order spectrum that is also more extended in the spatial
(cross-dispersion) direction.

Other potential sources of stray light in the \stis\ echelle modes
include scattered light in the telescope assembly, large angle dust
scattering, grating ghosts, reflections off of detector windows and
other instrument components along the light path, and perhaps others.
There is also a component of the background caused by particle
radiation and phosphorescent emission from the detector window, which
is particularly important for the NUV MAMA observations (Kimble \etal\
1998).

\section{Deriving the On-Order Background Spectrum}
\label{sec:algorithm}

The background in \stis\ echelle observations is quite complex, with
several sources of scattered and background light.  Unfortunately,
little information on the scattering properties of the instrument is
available in the astronomical literature or through the various \stis\
instrument teams.  We point the reader to the short discussion of the
background and scattered light properties of the instrument by Bowers
\etal\ (1998) and Landsman \& Bowers (1997) for the best summaries 
provided by the \stis\ IDT.

The level of scattered light can vary significantly within an order,
as judged by the residual flux levels in the saturated cores of strong
interstellar absorption lines in gross (non-background corrected)
spectra.  Thus a common flux offset to the data at all wavelengths is
insufficient as a background estimate.  The appropriate zero levels of
such saturated lines can vary by $\sim10\%$ of the local gross
continuum within a given order.  Furthermore, the background level
also varies significantly from order to order, making an average
global correction inappropriate.

Given the complexities of the origins of the scattered and background
light (and the instrument in general), we have developed an algorithm
for estimating the on-order background spectrum that does not
explicitly model the sources of background light.  Instead, we
empirically estimate the on-order background spectrum given the
structure in the inter-order regions of the MAMA detector for each
individual spectral order.  

Similar empirically-derived approaches have been successfully applied
for various other instruments using echelle gratings with
two-dimensional detectors (e.g., Churchill \& Allen 1995; Jenkins
\etal\ 1996).  Our method is different than those developed by
Churchill \& Allen (1995) for application to the Hamilton Echelle
Spectrograph and by Jenkins \etal\ (1996) for the IMAPS instrument.
In those cases the contribution to the on-order gross spectrum from
the adjacent orders is derived using the known cross-dispersion
profile of the closely-spaced spectral orders.  The effects of order
overlap are not as significant in the \stis\ echelle-mode observations
as in either of these instruments.  The background in our case is
dominated by the broad halo surrounding an order, with much smaller
contributions from the halos of the adjacent orders; echelle
scattering can also be important (see \S \ref{sec:stis}).

\subsection{The Method}
\label{subsec:method}

In this section we present the procedure we have developed for
deriving the on-order background for \stis\ E140H and E230H
observations.  Our approach is to fit the cross-dispersion profile of
the observed inter-order light in a manner that correctly estimates
the background at each position in the on-order spectrum.  The steps
used in our background removal procedure are as follows.  (1) Produce
two-dimensional rectified gross and error images of each order using
\calstis. (2) Compress the image for a given order along the
dispersion direction, thereby producing an average cross-dispersion
profile.  (3) Fit and subtract a rough background to the average
cross-dispersion profile derived in step 2.  (4) Identify the center
of the spectral trace in the background-subtracted average
cross-dispersion profile.  (5) Identify the inter-order background
regions in the gross image as those having less than 3.5\% of the peak
on-order flux.  (6) Smooth the inter-order background with a
two-dimensional Gaussian kernel. (7) Fit the cross-dispersion
inter-order background profile at each wavelength with a seventh-order
polynomial, weighting points nearer the spectral trace more heavily
than more distant points.  This step produces a two-dimensional
background image after the background has been fit for each wavelength
(e.g., Figure \ref{fig:twodimen}).  (8) Extract one-dimensional gross,
error, and background spectra from their respective two-dimensional
images by summing over a seven-pixel extraction box centered on the
spectral trace (summing in quadrature in the case of the error
spectrum).  (9) Smooth the background spectrum with a 17-pixel Lee
(1986) local statistics filter.  (10) Subtract the smoothed background
spectrum from the gross spectrum to produce the final, net spectrum.

The process presented above is relatively simple.  However, each of
the steps has been optimized to give the best background fit, and
therefore each requires a slightly more thorough discussion.  We
present the details of each of the steps of our technique below.

(1) For each order in the observation, we produce the
geometrically-rectified two-dimensional gross and error images, which
we denote \gij\ and \eij, respectively, using the standard \calstis\
software available at the Space Telescope Science Institute.  For our
purposes, $i$ and $j$ refer to the dispersion and cross-dispersion
direction, respectively.  In this case, increasing values of $i$ imply
increasing wavelengths, $\lambda$.  These images have been flux and
wavelength calibrated, though our method works just as well for count
rate images.

(2) We compress the gross image of a given order along the dispersion
direction, creating an order-averaged cross-dispersion profile, $a_j$:
\begin{equation}
  a_j \equiv \sum_{i} g_{i,j} .
\end{equation}
We use this average profile to determine the $y$-position of the
center of the spectral trace, $y_0$, and to define the background
regions of the rectified gross image.

(3) The array $a_j$ contains contributions from both the on-order
spectrum and the on-order and inter-order backgrounds.  To approximate
the average background pedestal in the collapsed spectrum, we fit a
fourth order polynomial to the cross-dispersion averaged profile,
excluding the inner 40\% of the array.  The rectified images are
slightly smaller than twice the separation between the order of
interest and its neighbors in the raw MAMA image.  (The inter-order
separation in the raw MAMA image scales linearly with the central
wavelength of the order.)  For short wavelength observations, the
fitting process described above excludes the central 12 pixels (the
FWHM of the spectral trace is $\sim2-3$ pixels).  This rough fit is
subtracted from $a_j$, producing a background-subtracted
order-averaged cross-dispersion profile, $a_j^0$.

(4) We identify the $y$-position of the center of the spectral trace,
$y_0$, with the maximum of $a_j^0$.  Although a Gaussian fit may be
more appropriate in theory, in practice we find identifying the center
of the spectral trace by its maximum is sufficient.

(5) After identifying the center of the trace, we define the
background regions, $y_b$, of the two-dimensional gross image to be
those points where $a_j^0 <0.035 \, {\rm max}(a_j^0)$.  Those points
that meet this criterion will be used in fitting the cross-dispersion
profile of the inter-order background light.  We have experimented
with many different cut-off criteria for defining the background
regions in the two-dimensional rectified images.  The resulting
background estimate can be somewhat sensitive to this parameter.  Our
experimentation suggests a cut-off at $3.5\%$ of the peak collapsed
spectrum intensity is appropriate in most cases.  The derived
backgrounds are similar if one uses cut-off levels in the range
$\sim2\%-5\%$.  The $y$-coordinates of the background regions are the
same for each wavelength within an order.

(6) After determining which regions of the image \gij\ make up the
off-order background, we smooth those regions with a two-dimensional
Gaussian having FWHM of 1.0 and 2.5 pixels in dispersion and
cross-dispersion directions, respectively.  This reduces some of the
statistical noise, making the fitting process less susceptible to
random noise fluctuations.

(7) Next we produce a background image, $b_{i,j}$, of the same
dimension as the gross and error images by fitting a one-dimensional
seventh-order polynomial to the cross-dispersion background profile at
each point $i$ (each wavelength $\lambda$).  Without the smoothing of
the previous step, the background image produced, $b_{i,j}$, contains
an excessive amount of small-scale structure.  The importance of each
cross-dispersion point, $y_j$, in the fitting process is weighted by
$|y_j-y_0|^{-1/2}$, so that regions closest to the on-order spectrum
are weighted more strongly than points at large distances from the
spectral trace.  Several different weighting schemes were tested, and
the results were relatively insensitive to the weighting scheme for
those weighting points toward the center more heavily.  More uniform
weighting schemes generally do not work as well.  Only those points
$y_j$ contained within $y_b$ are used in the fitting process.

We have found that fits using polynomials of higher order than seventh
often give spurious results (particularly given the small number of
background points used in the fits at short wavelengths).  Lower-order
fits often fail to adequately describe the curvature in the
inter-order light distribution as well as the adopted seventh-order
fits.

(8) After creating a background image, we extract one-dimensional
gross, background, and error spectra using a seven-pixel extraction
box.  For the gross and background images, the seven points centered
on the spectral trace are summed to produce the on-order gross
spectrum ($g_\lambda$) and initial background spectrum
($b_\lambda^\prime$):
\begin{equation}
	g_\lambda \equiv g_i = \sum_{j=y_0-3}^{y_0+3} g_{i,j}, 
\end{equation}
and
\begin{equation}
	b_\lambda^\prime \equiv b_i = \sum_{j=y_0-3}^{y_0+3} b_{i,j} .
\end{equation}
The error spectrum, $e_\lambda$, is produced by adding the seven
central pixels in quadrature:
\begin{equation}
	e_\lambda \equiv e_i = 
	\left( \sum_{j=y_0-3}^{y_0+3} e_{i,j}^2 \right)^{1/2}.
\end{equation}
We have chosen to adopt a seven-pixel summation extraction box to
match the standard \calstis\ routines, though other extraction box
sizes can be adopted.  An optimal extraction routine may also be used.

(9) The final background spectrum, $b_\lambda$, is produced by
smoothing the extracted background spectrum $b_\lambda^\prime$ with a
17-pixel Lee (1986) local statistics filter.  The Lee filter
effectively removes high-frequency noise in flat regions of the
spectrum while leaving high-contrast edges in the spectrum intact.
This is particularly important for spectra containing narrow
interstellar absorption lines.

(10) The final net spectrum, $n_\lambda$, is the difference between
the gross and smoothed background spectra, $n_\lambda \equiv g_\lambda
- b_\lambda$.

\subsection{Illustration of the Method}
\label{subsec:illustration}



\begin{figure}
\epsscale{0.9}
\plotone{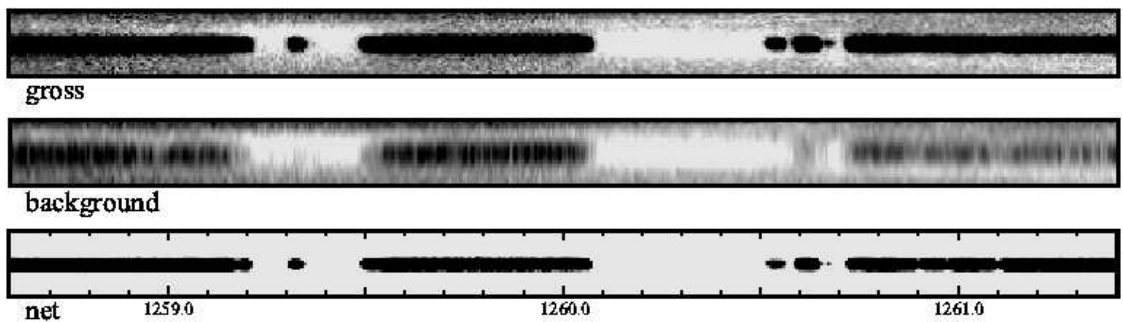}
\caption{Sections of the two-dimensional rectified gross (\protect\gij), 
background (\protect\bij), and net ($n_{i,j}$) images for STIS E140H
observations of the star HD~218915.  These images cover the wavelength
range 1258.6 to 1261.4 \AA\ in order 334.  A wavelength scale is given
on the bottom net spectrum image.  All images are shown on the same
intensity scale, which is truncated at zero flux.  In this example,
the background image has not been smoothed before subtracting it from
the gross image.  The average residual flux in the inter-order regions
of the net image is less than 0.09\% of the average on-order flux and
has a dispersion $\la 0.3\%$ of the average on-order flux.  A small
amount of echelle scattering of the saturated interstellar lines into
the inter-order background can be seen below the right-most portions
and above the left-most portions of the gross image.  The
high-frequency noise observable in the background image is removed
after extracting the one-dimensional background spectrum (see text).
\label{fig:twodimen}}
\end{figure}

\begin{figure}
\epsscale{0.9}
\plotone{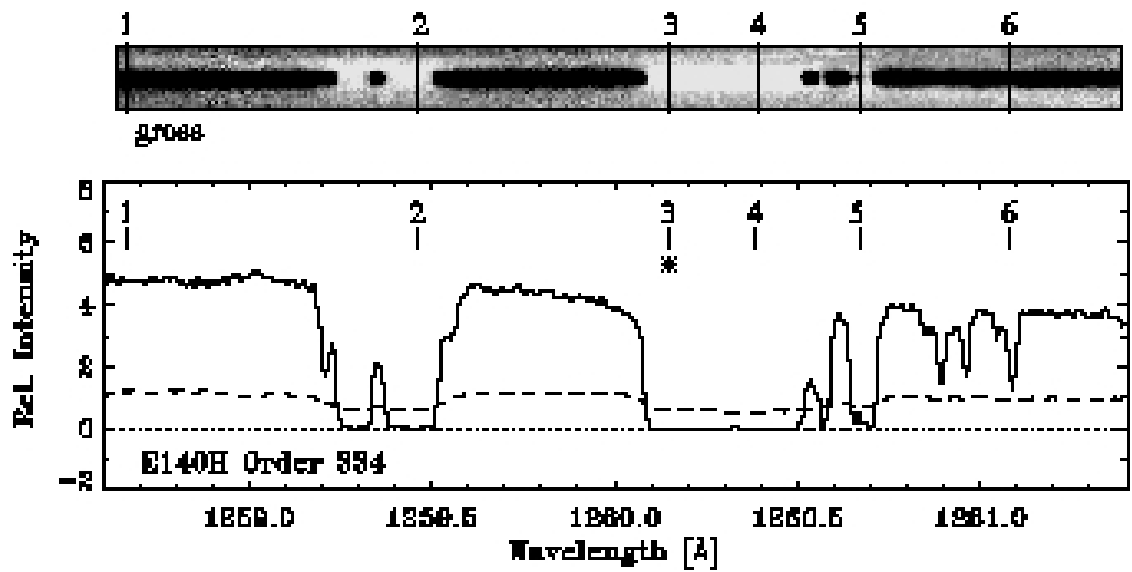}
\caption{The final extracted net spectrum ($n_\lambda$; solid histogram)
and background spectrum ($b_\lambda$; dashed line) for the spectral
region in Figure \protect\ref{fig:twodimen} shown with the gross
spectral image (\protect\gij).  Several positions in the spectrum are
marked, and the cross-dispersion profiles at these positions are shown
in Figure \protect\ref{fig:xdisp}.  The background at position 3 is
slightly over-estimated, leaving a negative residual flux in the short
wavelength portion of the saturated \protect\ion{Si}{2} absorption.
\label{fig:specimage}
}
\end{figure}

Figure \ref{fig:twodimen} shows a portion of the raw, two-dimensional
rectified gross image (\gij) of order 334 taken from the E140H
observation shown in Figure \ref{fig:fullmama}.  Also shown is the
corresponding two-dimensional background image ($b_{i,j}$) derived as
described above, and the difference of these images ($n_{i,j} \equiv
g_{i,j} - b_{i,j}$).  All of the images shown in Figure
\ref{fig:twodimen} are displayed with the same intensity scale, which
is truncated at a low level to show the structure of the inter-order
background.  The wavelength coverage of this section of the spectrum
is $\lambda \approx 1258.6$ to $1261.4$ \AA; this region includes
absorption from \SII\ \wave{1259} and \SiII\ \wave{1260} (as well as
weaker absorption from \FeII\ \wave{1260} and \ion{C}{1} \wave{1261}).
The final extracted net spectrum ($n_\lambda$) and background
($b_\lambda$) are shown in Figure \ref{fig:specimage}.

Several aspects of Figure \ref{fig:twodimen} are noteworthy.  First,
the echelle scattered images of the saturated interstellar absorption
lines are vaguely visible in the inter-order regions of the \gij\
image.  These can also be seen to some extent in the model background
image, \bij.  The background image contains the imprint of
echelle-scattered light since the fitting process does not
specifically exclude such regions.  Second, the background image,
\bij, shows significant high-frequency variations in the dispersion
direction.  This is a result of fitting each column (wavelength)
independently.  The application of the Lee filter to the extracted
one-dimensional background spectrum, $b_\lambda^\prime$, removes much
of this high-frequency noise from the final background spectrum,
$b_\lambda$ (see \S \ref{subsec:method}).  The inter-order regions of
the background-subtracted image, $n_{i,j}$, are relatively smooth and
very near zero residual intensity.  There is little residual
large-scale structure in the background regions of the $n_{i,j}$
image.

Figure \ref{fig:xdisp} shows several examples of cross-dispersion
profiles (histogram) extracted from the rectified two-dimensional
image of this order.  The cross-dispersion profiles are numbered, and
the positions of these profiles in wavelength space are marked on the
image and spectrum shown in Figure \ref{fig:specimage}.  These
profiles are drawn from several different regions of the spectrum,
including continuum regions (position 1) as well as regions that
include weak (positions 5 and 6) and strong (positions 2 through 4)
absorption.  The smooth curves plotted on top of the cross-dispersion
profiles show our one-dimensional polynomial fits to the inter-order
background, and the squares mark the data points used for deriving
those fits.  One can see that the inter-order backgrounds are well fit
at most of these positions.  A counter-example is position 3, which is
marked with a star in the spectrum shown in Figure
\ref{fig:specimage}.  In this case the background is over-estimated. 
The difficulty in fitting this position is caused by the effects of
echelle scattering; we will discuss this in more detail in \S
\ref{sec:examples}.

To show how the background we are fitting in these cross-dispersion
profiles relates to the larger-scale distribution of light in the
two-dimensional MAMA images, we present a cross-dispersion profile
through several orders in Figure \ref{fig:xorders}.  The distribution
shown here is drawn from the raw (gross) MAMA observation of \hd
218915.  The identifications of the orders are given along the top of
the figure.  This particular cross-dispersion cut passes through the
saturated troughs of two strong interstellar lines: \ion{N}{1}
\wave{1200.2} in order 351 and \ion{Si}{2} \wave{1193.3} in order 353.
Figure \ref{fig:xorders} shows a background fit for order 351 as the
solid line.  The points used in this background fit are marked with
filled squares, as in Figure \ref{fig:xdisp}.

One can see that while the orders are closely spaced, the excellent
image quality of \hst\ concentrates the on-order light into the
central few pixels.  Thus it is only the wings of the halos from
adjacent orders that overlap and not the cores of the on-order light
distribution.  This can be compared with the cross-dispersion profiles
shown in Churchill \& Allen (1995) and Jenkins \etal\ (1996), where
the overlap of light from adjacent orders is much more significant.

\subsection{Potential Limitations}
\label{subsec:limits}

The procedure we have outlined above is quite simple and can be
summarized as a single step: we fit the cross-dispersion profile of
the inter-order light to estimate the on-order background spectrum.
It seems established, however, that the cross-disperser gratings
provide little to the scattered light budget of the \stis\ echelle
modes (Bowers \etal\ 1998; Landsman \& Bowers 1997).  The more
important causes of scattered and background light in the echelle
modes are the PSF halo, the detector halo, echelle grating scattered
light, and the particle/phosphorescent background.  Neither the halos
nor the echelle scattering give a pure cross-dispersion profile of
scattered light.  It is important to consider the effects this may
have on the approach outlined in \S \ref{subsec:method}, given that we
are only fitting the cross-dispersion profile of the background light.

The PSF and detector halos provide (roughly) axisymmetric scattering
about a given point in the spectral trace (see lamp spectrum in Figure
\ref{fig:x2dcartoon}).  Thus, this type of scattering distributes
light both in the dispersion and cross-dispersion directions.  The
inter-order light at a given point contains halo-scattered light that
originates from several different wavelengths.  The weighting of data
points close to the spectral trace in our inter-order background fits
causes light scattered at small angles from the dispersion direction
by the PSF and detector halos to play an important role in the
resulting estimate of the on-order background spectrum.

The broad halos surrounding an order seem to be the dominant source of
on-order background.  For continuum sources the effects of the halos
of orders $m\pm1$ are significantly less than the halo of the order
$m$.  To quantify this we have derived a composite cross-dispersion
profile showing the distribution of halo light as a function of
distance from an order.  Figure \ref{fig:psf} shows composite
cross-dispersion profiles derived for two wavelength regions.  The top
plot shows a region near 1216 \AA\ derived from E140H observations of
strong Ly$\alpha$ emission the solar analog $\alpha$ Centauri A
(kindly made available to us by J. Valenti).  The bottom plot shows a
region near 2800 \AA\ derived from \ion{Mg}{2} emission from $\alpha$
Orionis.  The intensities have been normalized to unity at the center
of each order.  The points are averages of emission from two orders
(346 and 347 for the E140H data and 276 and 277 for the E230H data)
drawn from the brightest emission regions in these spectra.  We have
averaged only one side of each order to exclude the brightest
echelle-scattered light.  The filled circles have intensities $\leq
3.5\%$ of the peak and would hence be used in constructing the
background using our procedure (see
\S \ref{subsec:method}).  The bump of excess emission centered at $|
\Delta y | = 18$ (marked as $m\pm1$) in the E140H plot is caused by 
weak emission from the adjacent orders.  The adjacent orders in the
E230H plot are located at $| \Delta y | = 37$, i.e., beyond the edge
of the plot.

At the position of the adjacent order for the E140H data ($| \Delta y
| = 18$ in Figure \ref{fig:psf}), the expected relative intensity of
the broad halo from the central order is only a few percent of its
peak (extrapolating the halo contribution marked with filled circles
to $| \Delta y | = 0$).  The contribution from the halos of adjacent
orders to the order being extracted is therefore expected to be only a
few tenths of a percent of the peak on-order flux.  Similar limits can
be derived from the lack of strong spectral features from adjacent
orders $m\pm1$ in the gross spectrum of an order $m$.  The
contribution of the halos from orders $m\pm2$ to an order $m$ should
be less than 0.1\% of the contribution of its own halo.  For the E230H
distribution, the halos have a flatter decline with distance from the
center of the order.  However, the overall level of the halo is
significantly lower relative to the peak intensity in the center of
the order compared with the E140H distribution.  This is consistent
with the much lower level of the backgrounds seen at 2800 \AA\ than at
\lya\ (see \S \ref{subsec:fraction}).

For continuum regions the halo contribution from the current order is
much more important than that from the adjacent orders.  This is in
part why we have not adopted a method similar to that of Churchill \&
Allen (1995) or Jenkins \etal\ (1996).  In cases where strong emission
lines are present in an adjacent order, it is true that the
contribution from the halo of an adjacent order $m\pm1$ can be more
important than the local halo if the emission from the local order $m$
is very weak.

Light scattered by the echelle gratings is the single most important
limiting factor in our approach.  In theory, if the spectrum scattered
into the inter-order region is smooth over relatively large scales,
echelle-scattered light could be accurately fit by our technique.  In
that case, the inter-order light is a smoothly varying function in the
cross-dispersion direction, i.e., there is a degeneracy between
cross-disperser and echelle scattering in the inter-order regions for
a smooth spectral source.  However, there are small-scale structures
in the scattered spectrum (e.g., interstellar absorption features,
particularly the strongly saturated lines).  In the case where strong
or numerous small-scale spectral features are scattered by the echelle
gratings into the inter-order regions, the validity of our approach
may break down.  Echelle scattering of local interstellar features
into the inter-order regions can bias the cross-dispersion fits (see
\S \ref{sec:examples}).  This will be particularly true if the
scattering by the echelle gratings makes up a very large fraction of
the scattered light budget.

In general, we believe the relative echelle scattering contribution to
the total on-order background spectrum is less important than the
detector halo (in terms of the relative intensities).  Using the E230H
lamp observation shown in Figure \ref{fig:x2dcartoon}, we estimate
that the echelle scattering contribution to the interorder background
at $\lambda \sim 2450$ \AA\ is $<5\%$ of the contribution from the
halo seen surrounding strong emission lines.  Using the aforementioned
E140H observations of \lya\ emission from $\alpha$ Cen, we have found
that the peak intensity of the echelle-scattered emission is
equivalent to the strength of the halo shown in Figure
\ref{fig:psf} at $| \Delta y | \approx 10$ pixels (i.e., 
$\la10\%$ of the expected halo intensity near the center of the
order).

However, the impact of echelle scattering on our method is more
important than its intensity relative to the broad detector halo would
suggest.  Discrete features in the echelle-scattered inter-order light
adversely effect the cross-dispersion fits in some cases.  Thus the
results for regions near very strong, sharp absorption (or emission)
lines will need to be viewed with caution, particularly at short
wavelengths where the inter-order separation is smaller.  This is
discussed more fully in \S
\ref{sec:examples}.

\section{Results}
\label{sec:examples}

\subsection{Examples and Uncertainties}

In this section we show several examples of \stis\ high-resolution
echelle data extracted as discussed in \S \ref{sec:algorithm}.  The
wavelength regions shown contain strongly-saturated interstellar
lines.  These flat-bottomed lines, which should have zero residual
flux, offer a chance to assess the quality of the background
subtraction through their shape and flux level.  A detailed comparison
of \stis\ data extracted with our algorithm and archival \ghrs\ data
will be given in \S \ref{sec:ghrs}.  We draw examples in this section
both from archival \stis\ data and our own guest observer program
(STScI program \#7270).

Figures \ref{fig:218915}, \ref{fig:185418}, and \ref{fig:303308} show
examples of short wavelength ($\lambda < 1350$ \AA) E140H spectra for
the stars \hd 218915, \hd 185418, and \hd 303308, respectively.  The
four spectral regions shown in these figures contain saturated
absorption lines near 1200 \AA\ (the \ion{N}{1} triplet), 1260 \AA\
(\SII\ \wave{1259} and \SiII\ \wave{1260}), 1302 \AA\ (\OI\
\wave{1302} and \SiII\ \wave{1304}), and 1334 \AA\ (\CII\ \wave{1334}
and \CII$^*$ \wave{1335}).  The solid lines show the
background-corrected on-order spectrum, $n_\lambda$, while the dashed
lines show the on-order background spectrum, $b_\lambda$, that was
subtracted from the extracted gross spectrum.  For the region near
\CII\ \wave{1334}, we have co-added two spectral orders (315 and 316)
weighted by their respective error spectra.

Figure \ref{fig:303308long} shows several longer-wavelength spectra of
\hd 303308, including three spectral regions observed with the E230H
grating.  This figure shows four spectral regions that include the
saturated absorption lines \ion{Al}{2} \wave{1670}, \ion{Fe}{2}
\wave{2382} and \wave{2600}, and \ion{Mg}{2} \wave{2796}.  As in the
previous figures, the on-order background spectrum is shown as the
dashed line.

The representative spectra shown in Figures
\ref{fig:218915}--\ref{fig:303308long} exhibit several common features
of our data extraction routine.  First, we find that the ratio of the
on-order background spectrum to the on-order net (or gross) spectrum
is higher in the shorter wavelength observations.  This is consistent
with expectation, particularly given the smaller order spacing at
shorter wavelengths.  The ratio of on-order background to gross
spectrum will be discussed in more detail in \S \ref{subsec:fraction}.

Next, it is clear from Figures \ref{fig:218915} through
\ref{fig:303308long} that the quality of the background subtraction is
much better at longer than at shorter wavelengths.  For all the stars
shown in these figures, the absorption due to \CII\ \wave{1334} is
flat-bottomed at zero flux level.  Towards progressively shorter
wavelengths, the saturated absorption profiles tend to show excursions
from true zero with slight artifacts appearing in the black troughs.
Figure \ref{fig:218915blowup} shows expanded views of regions of the
\hd 218915 spectra containing strongly-saturated interstellar
absorption.  The profile of \SiII\ \wave{1260} in Figure
\ref{fig:218915blowup} shows a slight over-subtraction of 
the background on the lower-wavelength end of the otherwise black
absorption trough.  The \NI\ \wave{1199} profile seen in Figure
\ref{fig:218915blowup} shows over-subtracted regions at various points 
along the absorption trough.  The background in this case shows
significant spurious structure (e.g., near $\lambda = 1199.35$ \AA).
Over-subtraction of the background is seen for all three members of
the \NI\ triplet near 1200 \AA\ (see Figure \ref{fig:218915}).
However, \CII\ \wave{1334} shows only a very small over-subtraction of
the background.  The long wavelength observations of \hd 303308 shown
in Figure \ref{fig:303308long} show no such artifacts, and the
saturated absorption profiles are at the correct zero level over their
entire breadth.

\begin{figure}
\epsscale{0.8}
\plotone{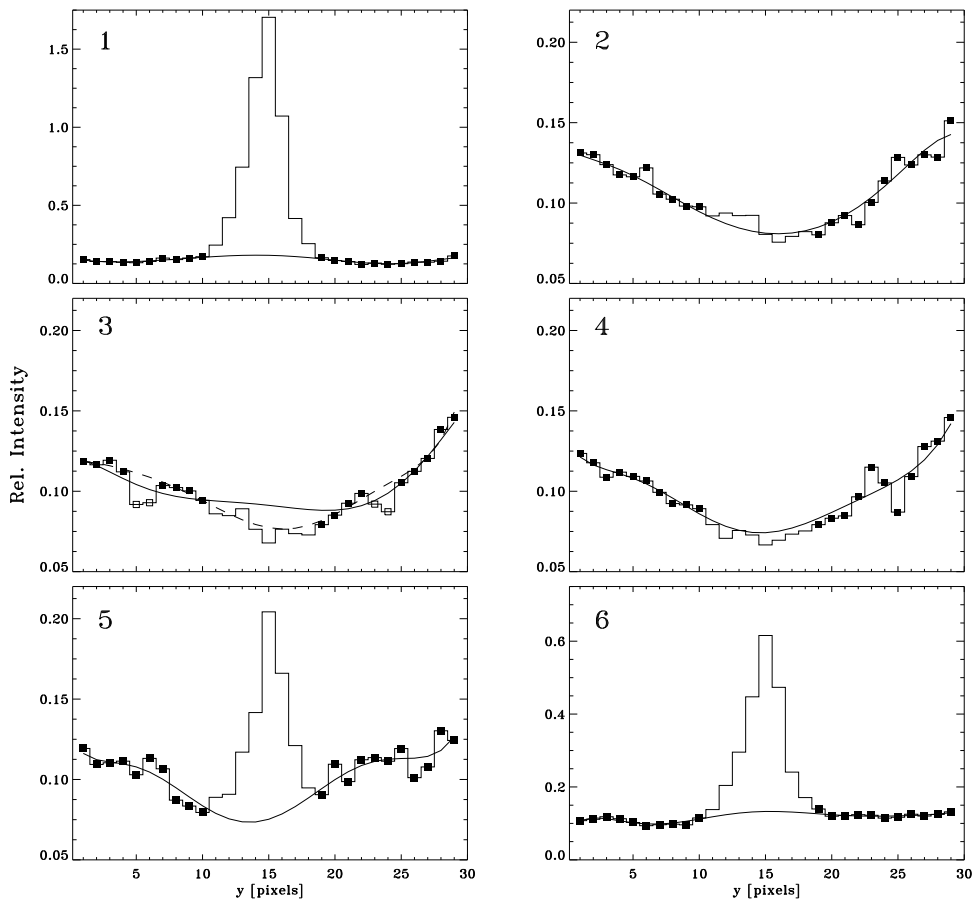}
\caption{Sample polynomial fits to the cross-dispersion background light 
profiles in the E140H observations of HD~218915.  These panels show
examples of the cross-dispersion polynomial fits for six wavelengths
designated in Figure \protect\ref{fig:specimage}.  The histograms show
the cross-dispersion profiles, while the solid lines show the fits to
the background.  The squares mark the points used in the fits.  The
background fits shown at positions 2 and 4 are examples of good fits
in the cores of strong lines.  The background fit to the
cross-dispersion profile in region 3, a position in the absorption
trough of \protect\ion{Si}{2} $\lambda1260$ (marked with a star over
the spectrum in Figure \protect\ref{fig:specimage}), over-estimates
the true background in this position.  This is caused by the echelle
scattering of other regions of this absorption line into the
inter-order regions.  Pixels in the cross-dispersion profile that are
affected by this echelle scattering are marked with open squares in
panel 3.  The dashed line shows the results of fitting the same
cross-dispersion profile excluding these pixels.  In this case the
background fit is excellent.
\label{fig:xdisp}}
\end{figure}

\begin{figure}
\epsscale{1.0}
\plotone{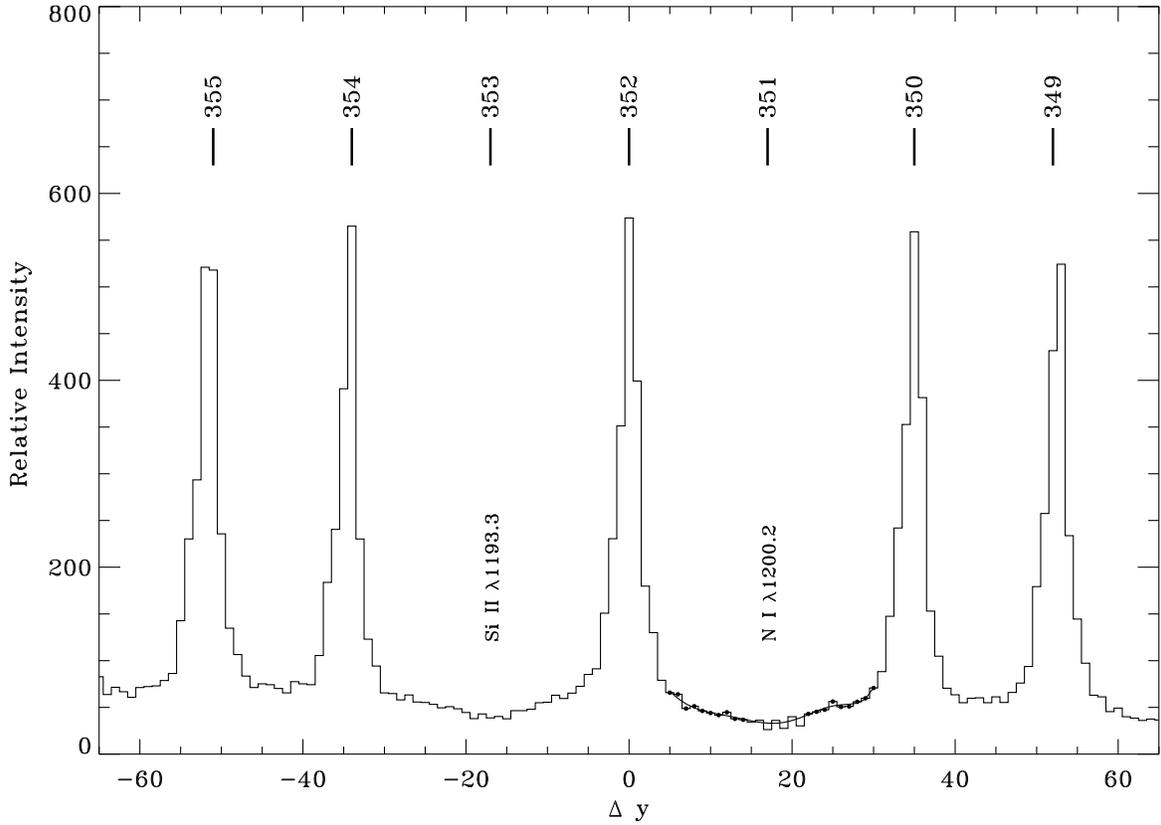}
\caption{A cross-dispersion view of  several spectral orders from the 
raw MAMA image, shown as the relative intensity versus pixels from
order 352.  The orders numbers are given above their expected
position.  This particular cut through the MAMA image is an average of
five pixels in the dispersion direction, and it samples two strongly
saturated absorption lines: \protect\ion{N}{1} $\lambda 1200.2$ in
order 351 and \protect\ion{Si}{2} $\lambda 1193.3$ in order 353.  We
show a sample cross-dispersion background fit for order 351 as the
solid line.  The points used in deriving this fit are marked with
solid squares.
\label{fig:xorders}}
\end{figure}

\begin{figure}
\epsscale{1.0}
\plotone{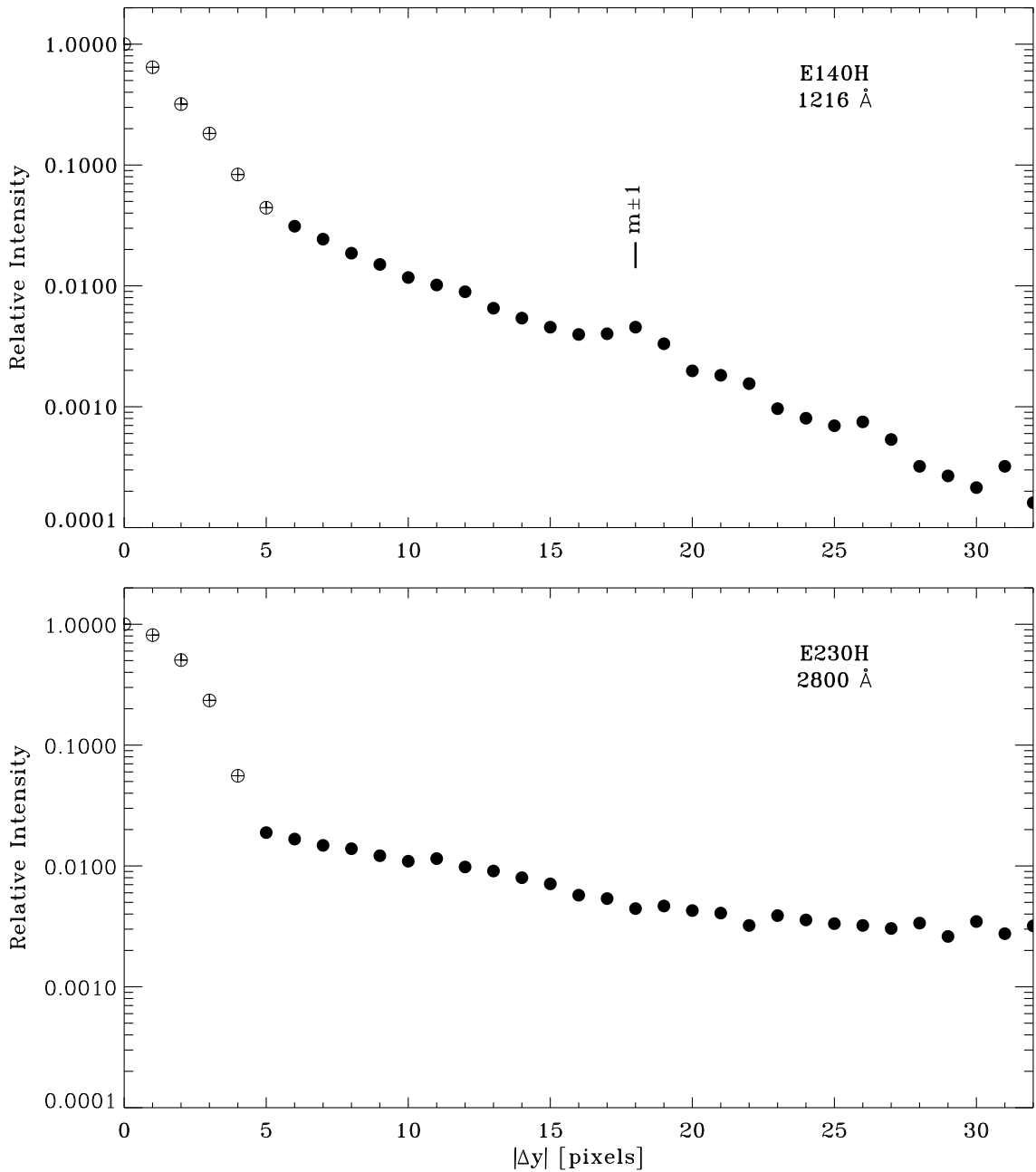}
\caption{Composite cross-dispersion profiles for a region near Ly$\alpha$ 
taken from E140H observations of the solar analog $\alpha$ Centauri A
(kindly made available to us by J. Valenti) and for E230H observations
of \protect\ion{Mg}{2} emission from $\alpha$ Ori (top and bottom,
respectively).  The intensity has been normalized to unity at the
center of the order.  The points for the E140H observations are
averages of two 15-pixel wide cross-dispersion profiles drawn from
orders 346 and 347.  We have averaged only one side of each order in
producing this plot to avoid the brightest regions of
echelle-scattered Ly$\alpha$ emission.  The bump of excess emission
centered at $| \Delta y | = 18$ (marked as $m\pm1$) is caused by weak
emission from the adjacent order.  The points for the E230H
observations were derived using three 20-pixel wide cross-dispersion
profiles from orders 276 and 275.  The adjacent orders are located at
$| \Delta y | = 37$ in this case and have less than 3\% of the flux in
the peak of the orders shown.  The filled circles have intensities
$\leq 3.5\%$ of the peak and would hence be used in constructing the
background using our procedure (see \S \protect\ref{subsec:method}).
The crossed circles denote points with intensities $> 3.5\%$ of the
peak.
\label{fig:psf}}
\end{figure}

\begin{figure}
\epsscale{0.9}
\plotone{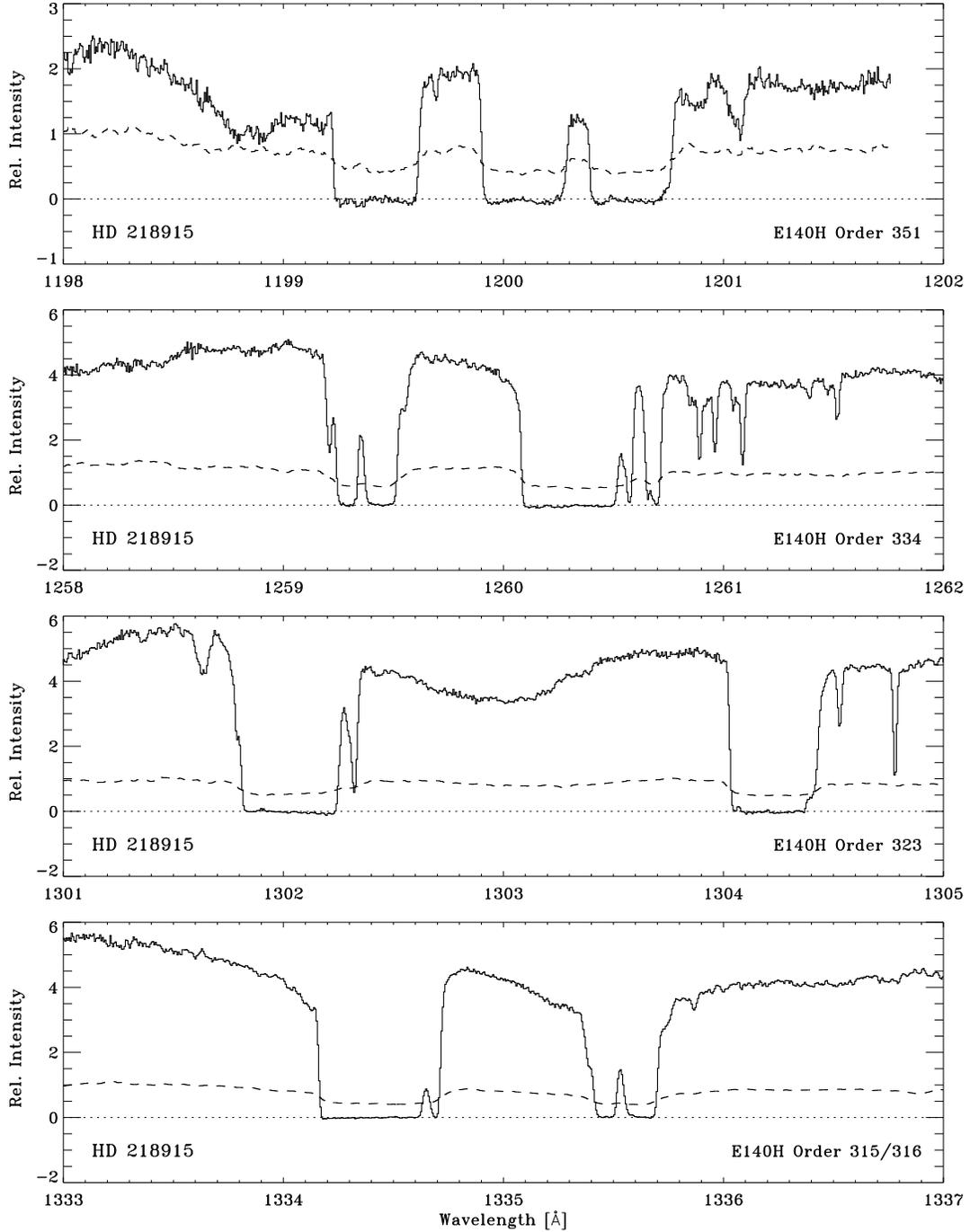}
\caption{Background-subtracted STIS E140H observations of the star
HD~218915.  Several strongly saturated interstellar lines are present
in these wavelength regions, including the \protect\ion{N}{1}
$\lambda1200$ triplet, \protect\ion{S}{2} $\lambda1259$,
\protect\ion{Si}{2} $\lambda1260$, \protect\ion{O}{1} $\lambda1302$,
\protect\ion{Si}{2} $\lambda1304$, \protect\ion{C}{2}
$\lambda1334$ and \protect\ion{C}{2}$^*$ $\lambda 1335$.  The
background that was subtracted from the gross data is shown as the
dashed line.
\label{fig:218915}}
\end{figure}

\begin{figure}
\epsscale{0.9}
\plotone{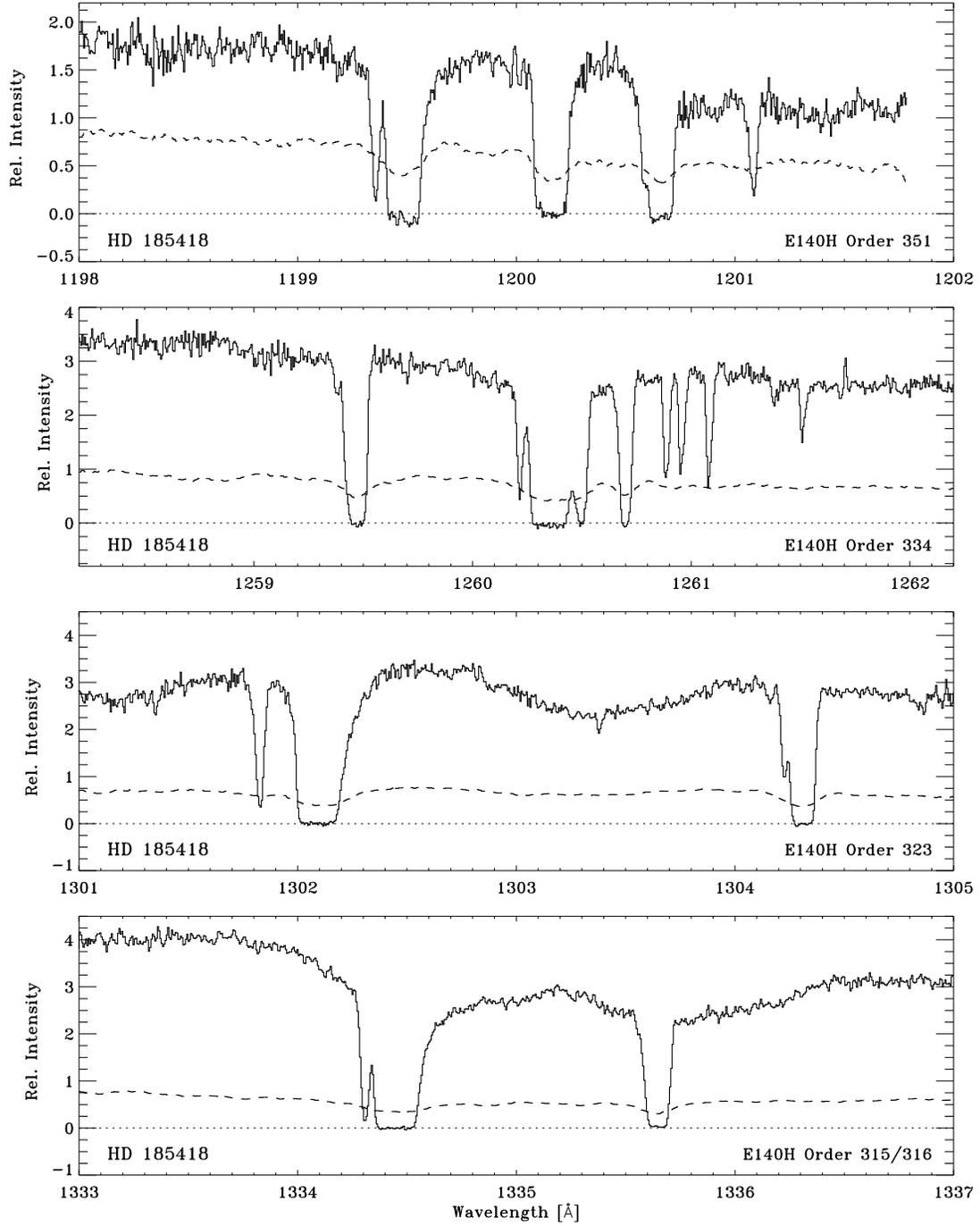}
\caption{As Figure \protect\ref{fig:218915}, but for the star HD~185418.  
These archival data were obtained as part of the ISM SNAP Survey
(Lauroesch \& Meyer 1999).
\label{fig:185418}}
\end{figure}

\begin{figure}
\epsscale{0.9}
\plotone{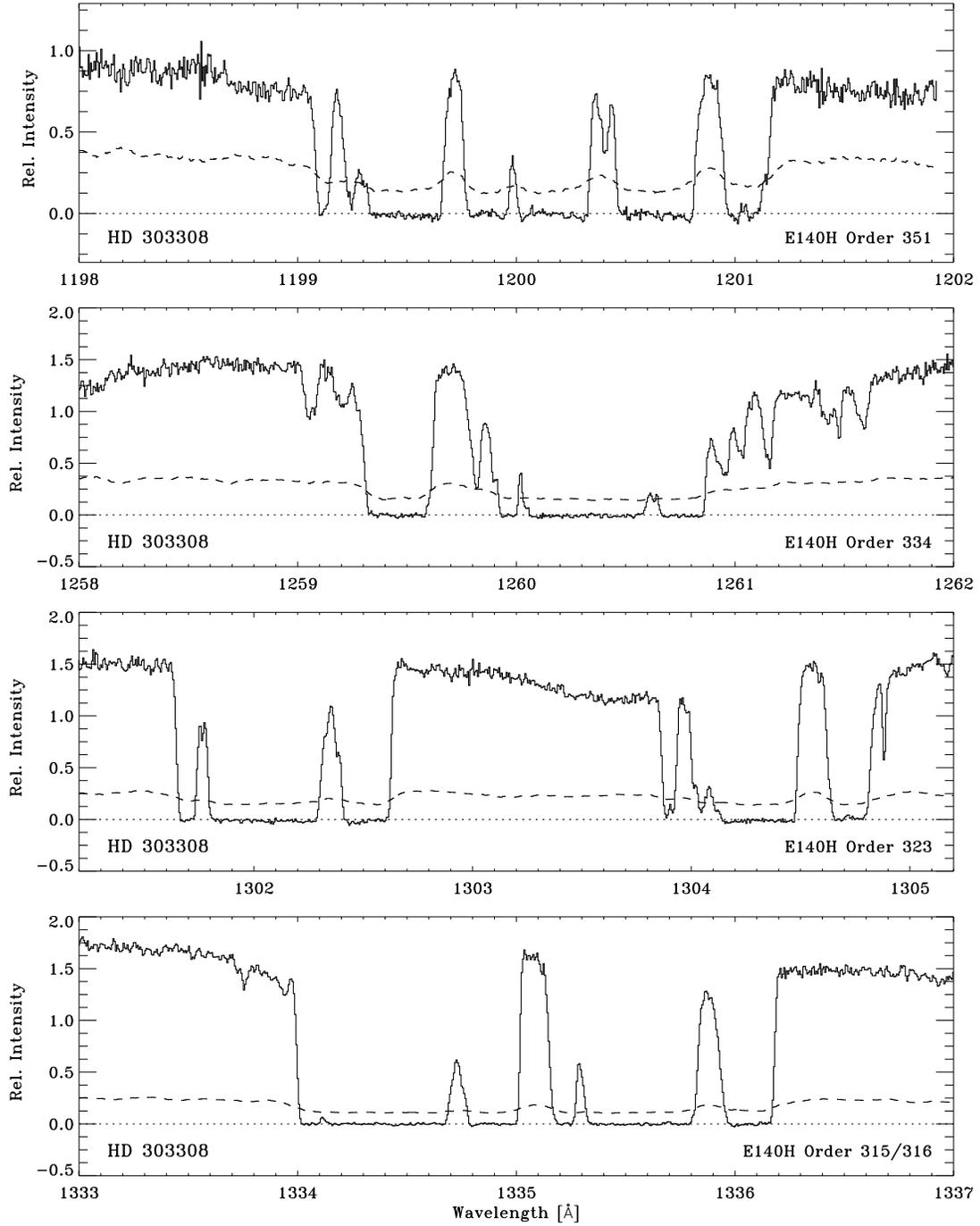}
\caption{As Figure \protect\ref{fig:218915}, but of the star 
HD~303308 in the direction of the Carina Nebula.  Several absorbing
components at different velocities are present in each line over the
wavelength interval shown.
\label{fig:303308}}
\end{figure}

\begin{figure}
\epsscale{0.9}
\plotone{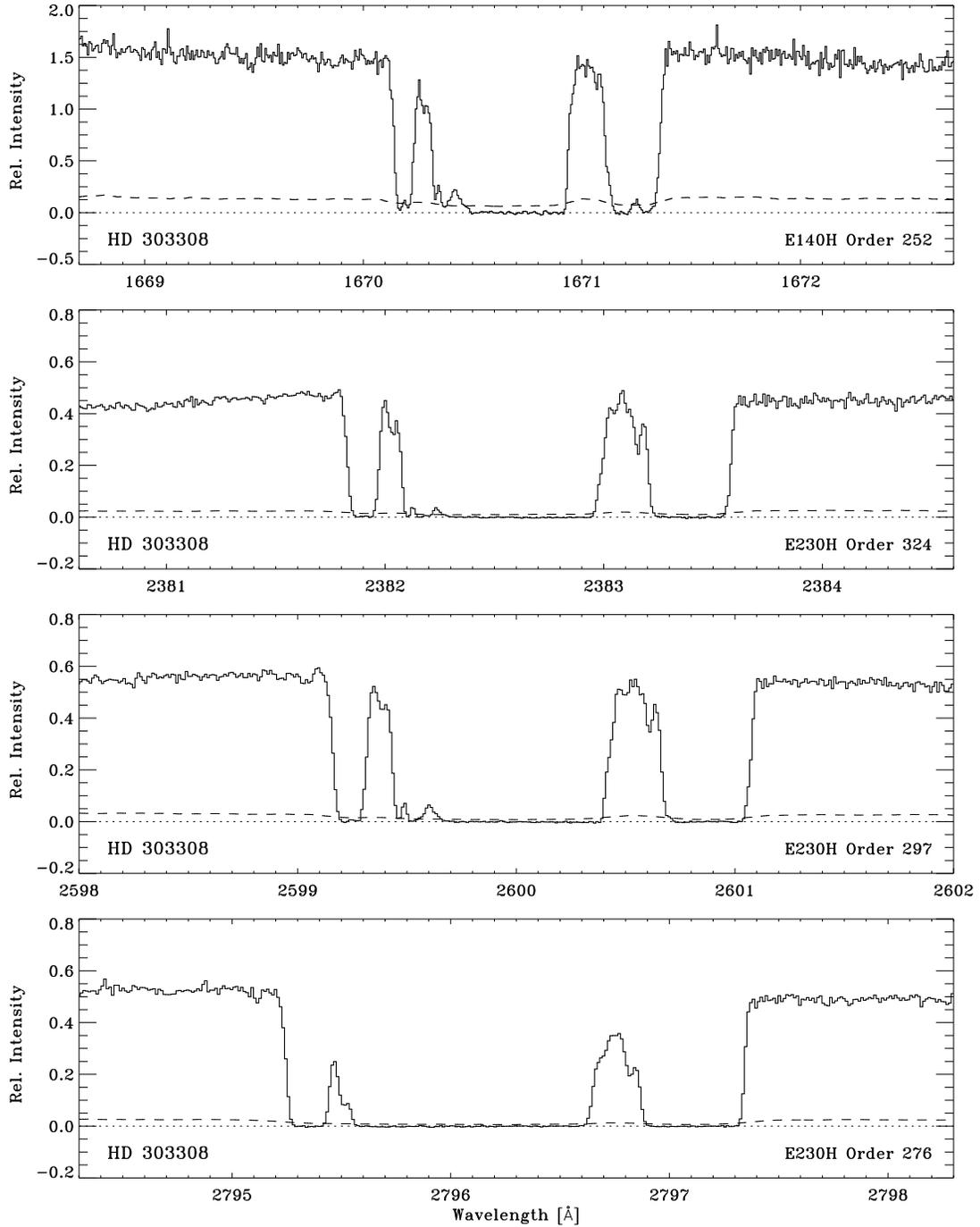}
\caption{As Figure \protect\ref{fig:303308}, but for longer wavelength 
data, including E230H observations.  The quality of the background
subtraction at the wavelengths shown here appears to be excellent.
The saturated absorption lines seen in this figure are (from top to
bottom): \protect\ion{Al}{2} $\lambda1670$, \protect\ion{Fe}{2}
$\lambda2382$, \protect\ion{Fe}{2} $\lambda2600$, and
\protect\ion{Mg}{2} $\lambda2796$.  Several absorbing
components at different velocities are present in each line over the
wavelength interval shown.
\label{fig:303308long}}
\end{figure}

\begin{figure}
\epsscale{0.5}
\plotone{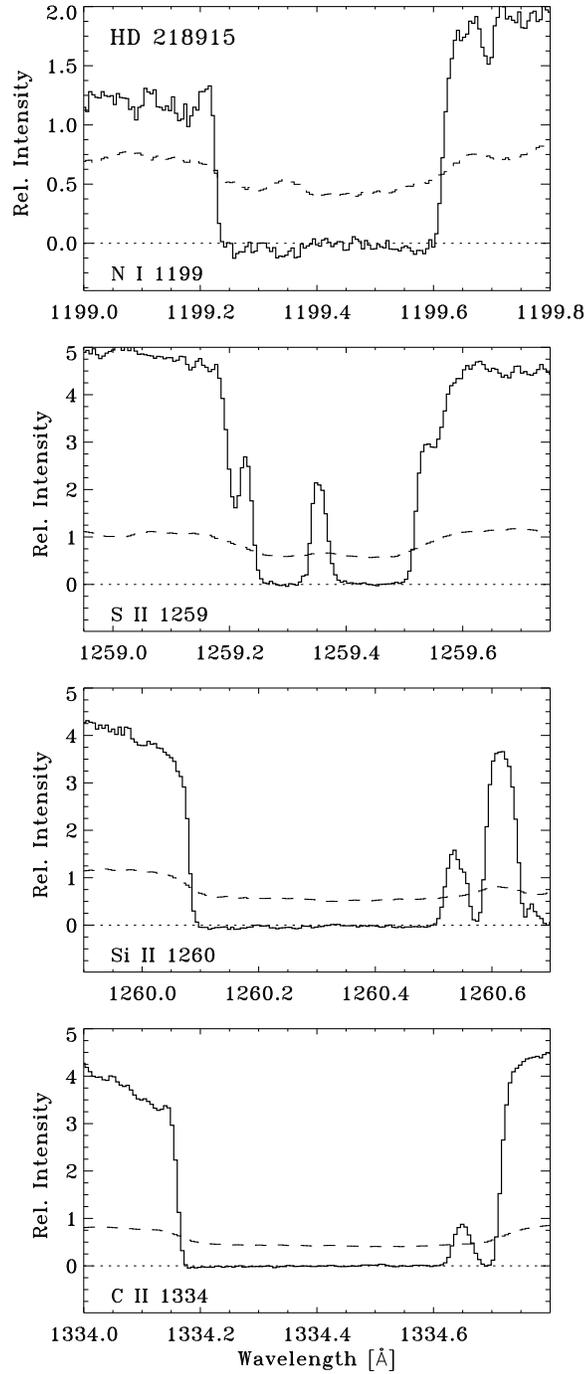}
\caption{Expanded views of several saturated interstellar lines 
towards \protect\hd 218915 (see Figure \protect\ref{fig:218915}).
This figure shows some of the background-subtraction artifacts that
are present in strongly saturated interstellar lines (e.g., the
structure present in the \protect\NI\ \protect\wave{1199} and
\protect\SiII\ \protect\wave{1260} profiles).
\label{fig:218915blowup}}
\end{figure}

This characteristic over-subtraction of the background in short
wavelength observations seems to be caused by the influence of echelle
scattering of strong lines into the inter-order background on our
cross-dispersion polynomial fits (see \ref{subsec:limits}).  Figure
\ref{fig:xdisp}  shows several examples of cross-dispersion profiles
extracted from the rectified two-dimensional image of order 334 in our
E140H observations of \hd 218915, as discussed in \S
\ref{subsec:illustration}.   Most of the cross-dispersion profiles 
shown in Figure \ref{fig:xdisp} are examples of regions where our
fitting procedure works well.  Position 3, marked with a star in the
spectrum shown in Figure \ref{fig:specimage}, shows a wavelength
region where our background fit overestimates the true background
level.  In this case, the saturated line core is over-subtracted by
$\sim 1\%$ of the local continuum.  The cross-dispersion profile for
position 3 contains two regions, marked with open squares in Figure
\ref{fig:xdisp}, with lower intensity than their surroundings (the
order is centered at $y=15$).  These discrete regions on either side
of the spectrum are images of strong interstellar absorption from the
on-order spectrum that have been scattered into the inter-order region
by the echelle grating.

The dashed line over-plotted on the cross-dispersion profile of
position 3 shows a polynomial derived by excluding these four points
from the fit.  The fit shown by the dashed line is significantly
better than that used in deriving the spectrum shown near the bottom
of the figure.  However, we do not at present have an algorithm to
identify and exclude such points from the cross-dispersion fits.  If
the spatial distribution of the echelle-scattered background were
derived, and the effects of this scattering removed from the
two-dimensional images before the cross-dispersion fitting described
here, our approach would likely have fewer problems at short
wavelengths.  The method developed by Churchill \& Allen (1995) for
use with the Hamilton Echelle Spectrograph does in effect build a
model two-dimensional background surface which is subtracted from the
raw data before standard spectra-extraction techniques are applied.
The derivation of the echelle scattering properties is, however,
beyond the scope of this work.

\begin{figure}
\epsscale{1.0}
\plotone{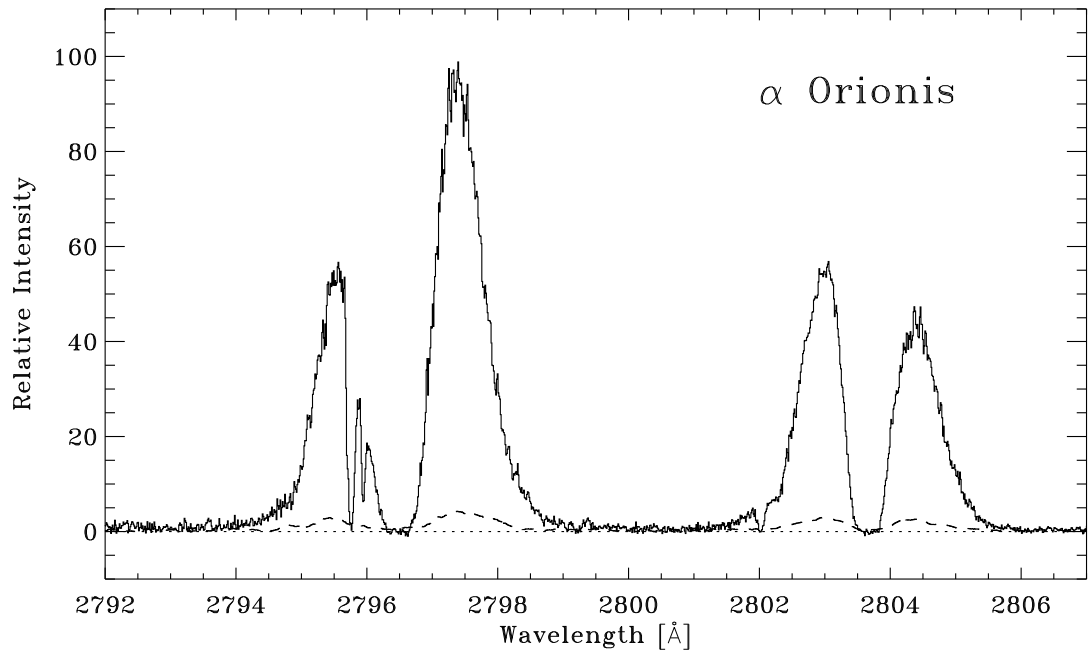}
\caption{An E230H spectrum of $\alpha$ Ori (Betelgeuse) in the wavelength
region covering the \protect\ion{Mg}{2} \protect\wave{2800} doublet.
The deep interstellar absorption is flat bottomed at zero flux,
showing our correction algorithm reproduces the correct zero point
even for late-type stellar spectra.
\label{fig:alphaori}}
\end{figure}

Our emphasis in presenting (and testing) our background subtraction
routines has been on observations of early-type stars, since these are
the most important for studying the interstellar medium.  However, our
reduction techniques should also be suitable for most other types of
observations.  To demonstrate the applicability of our approach to
other types of observations, we have reduced archival \stis\ E230H
observations of $\alpha$ Orionis (Betelgeuse).  The region surrounding
the \ion{Mg}{2} \wave{2800} doublet is shown in Figure
\ref{fig:alphaori}.  The saturated interstellar absorption lines seen
on top of the stellar emission lines are flat bottomed at zero flux,
as expected.  We have also tested our routine on the aforementioned
FUV \stis\ observations of the solar analog $\alpha$ Cen.  In general
our routines did an excellent job deriving the background for this
emission line dataset.  However, in the orders adjacent to the
extremely bright \lya\ emission from this star, the wings of the
detector halo extended well into the very weak emission from the
adjacent orders (see Figure \ref{fig:psf}).  This caused our routines
to over-subtract the very bright background from the very weak
continuum and line emission in the orders adjacent to \lya.  This is
the one instance we have identified where our procedures are not able
to appropriately fit the interorder background.  Our tests suggest
that, with this one exception, our extraction procedures are doing
reasonably well even for late-type stars.

\subsection{Caveats}
\label{subsec:caveats}

Given the difficulties in fitting the cross-dispersion profiles at
short wavelengths, where the halos of the closely-packed spectral
orders may overlap significantly and the echelle scattering affects a
greater fraction of inter-order pixels, we believe particular
attention should be paid to background uncertainties.  Lines having
residual fluxes less than $10-20\%$ of the local continuum flux should
be treated with care.  Astronomers working with such lines in
high-resolution \stis\ data should consider the effects of zero-point
uncertainties that depend on wavelength in deriving column densities
and other parameters.  The fluctuations in our background estimates
can be used to approximate the uncertainties in the overall zero level
for individual spectral regions.  For example, the standard deviation
about the mean of the background spectrum in the spectral region
encompassing the saturated \SiII\ 1260 \AA\ line is $\sim 0.6\%$ of
the local continuum.  The $2-3 \, \sigma$ fluctuations are a good
estimate of the potential error in the background spectrum in this
case (as judged by the maximum oversubtraction of the \SiII\ line).
For lines with residual fluxes that are 10\% of the local continuum,
this implies potential systematic errors in the optical depth of order
10\%.

Although the background subtraction at short wavelengths, e.g.,
$\lambda \la 1300$ \AA, often shows evidence of slight artifacts in
the saturated absorption troughs of deep lines, we find little
evidence for such problems in the longer wavelength observations.
This is due to the larger order separation, generally higher
signal-to-noise ratios, and smaller contribution of the on-order
background to the gross spectrum.  Data taken with the E140H grating
at wavelengths in excess of $\lambda \ga 1600$ \AA\ and with the E230H
grating at wavelengths $\lambda \ga 2200$ \AA\ (where
$b_\lambda/g_\lambda \la 0.1$; see \S \ref{subsec:fraction}) are
likely more secure than shorter wavelength data in each of the
gratings.  And while we correctly state that these artifacts are
present for wavelengths shorter than 1300 \AA, the difficulties are
most apparent wavelengths shortward of \lya.  However, analysis of
detailed line shapes for strong absorption should still be approached
with caution for $\lambda \la 1300$.

In general astronomers using a routine like the one outlined here may
decide to apply a second-order correction that adjusts the output
spectrum by a simple vertical flux offset.  Such a shift is designed
to bring the majority of a saturated line profile to the correct zero
level in a given order.  We have calculated the shift required to make
the average of the saturated line profiles of the short wavelength
lines observed towards \hd 303308 equal zero flux.  We find an offset
of $1-2\%$ of the local continuum is most appropriate.  Although we
are skeptical that this is the best way to correct spectra where the
saturated lines fall somewhat below zero, the correction is typically
modest.

The quality of the background estimation using the algorithm presented
here is dependent on the signal-to-noise ratio of the dataset.  In low
signal-to-noise ratio data, the polynomial fitting routine used in our
method can produce spurious background excursions.  Thus, if one wants
to accurately remove the on-order background using our procedures, it
is important to design observing programs that produce enough
inter-order light to be accurately fit.  For practical purposes, this
means producing datasets with signal-to-noise ratios in excess of 20
is important for accurate background determinations.

\subsection{The On-Order Background Fraction}
\label{subsec:fraction}

\begin{figure}
\epsscale{1.0}
\plotone{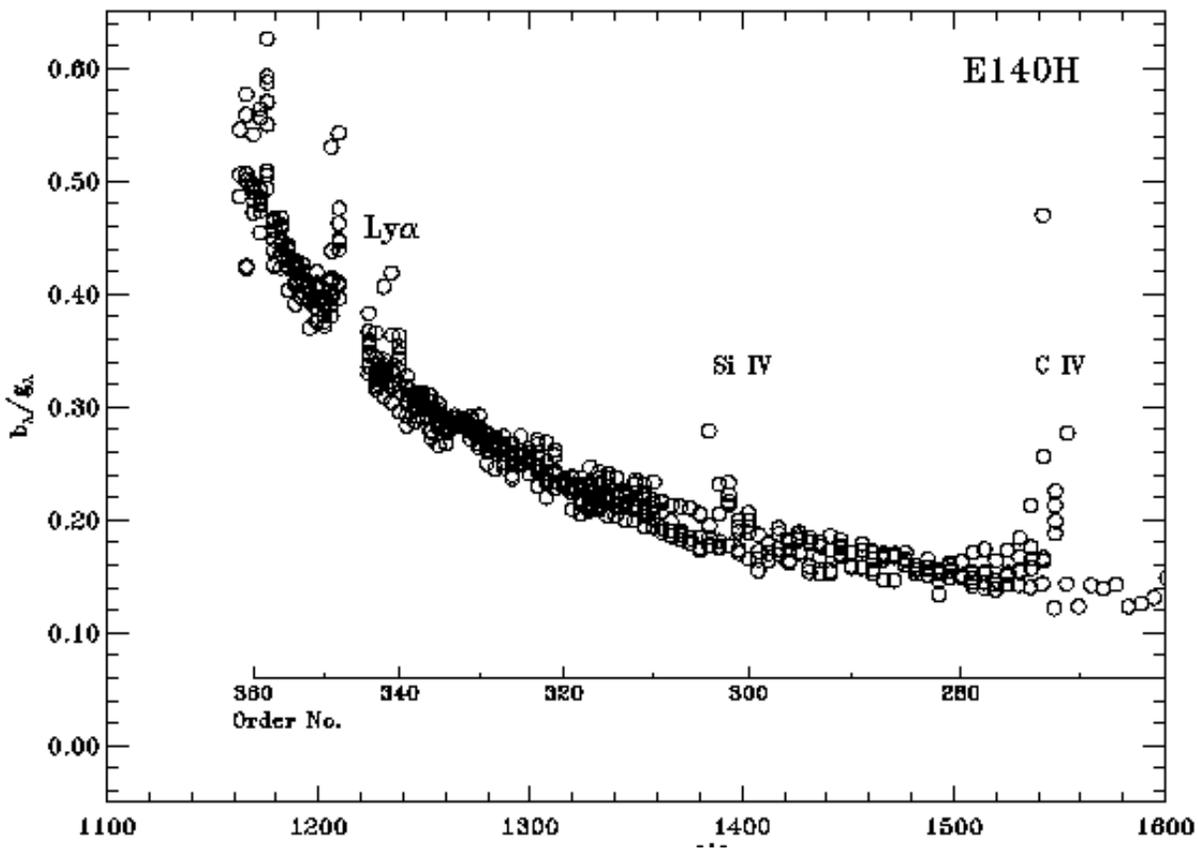}
\caption{The fractional contribution of the on-order background light,
$b_\lambda$, to the gross spectrum, $g_\lambda$ for several E140H
observations of early type stars.  Each point represents the median
value of $b_\lambda/g_\lambda$ for an individual order of a particular
observation.  Data are presented for nine sightlines taken from the
STIS archive or our own GO program; several sightlines were observed
in multiple wavelength regions.  Average and median
$b_\lambda/g_\lambda$ values for specific orders are given in Table
\protect\ref{tab:e140hfrac}.  Data taken through both the
$0\farcs2\times0\farcs09$ and $0\farcs2\times0\farcs2$ apertures are
shown.  The dispersion at certain wavelengths is very large due to the
presence of strong P~Cygni-like stellar wind profiles in several of
the stars (e.g., near \protect\ion{C}{4} $\lambda1550$) or strong
interstellar absorption (e.g., near Ly$\alpha$).  We have marked
several such regions in this figure.  We have excluded orders that are
dominated by interstellar Ly$\alpha$ absorption.  The ratio
$b_\lambda/g_\lambda$ decreases approximately as $\lambda^{-3}$.
These data were derived using observations of early-type stars.  The
$b_\lambda/g_\lambda$ ratio could depend on the spectral type.
\label{fig:e140hfrac}}
\end{figure}

\begin{figure}
\epsscale{1.0}
\plotone{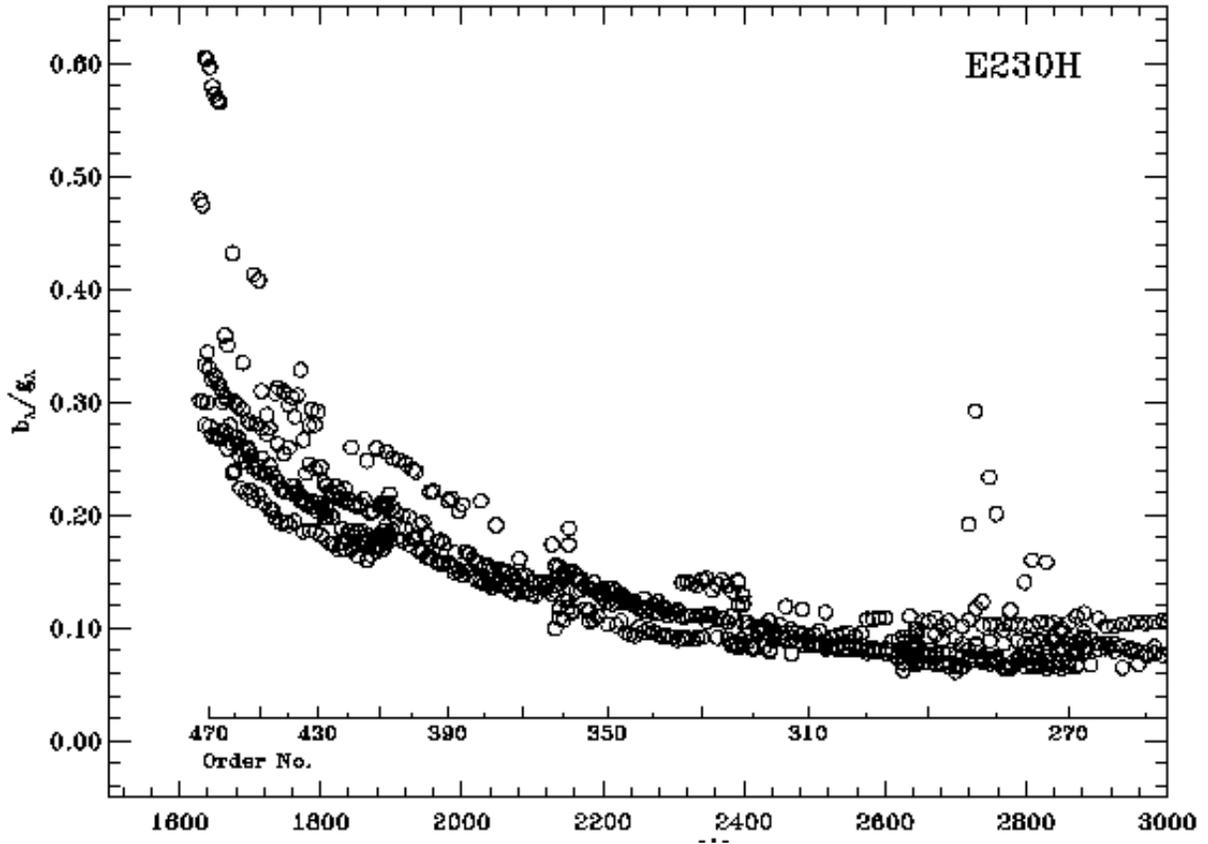}
\caption{As Figure \protect\ref{fig:e140hfrac}, but for data taken
with the E230H grating.  Data taken from 11 early-type stars, some of
which are observed in multiple wavelength regions, are shown.  The
ratio $b_\lambda/g_\lambda$ again decreases approximately as
$\lambda^{-3}$.  Average and median $b_\lambda/g_\lambda$ values for
specific orders are given in Table
\protect\ref{tab:e230hfrac}.
\label{fig:e230hfrac}}
\end{figure}

We have calculated the fractional contribution of the on-order
background light to the on-order gross spectrum as a function of
wavelength (spectral order) for a number of archival \stis\ datasets
as well as our own guest observer program data.  Figure
\ref{fig:e140hfrac} shows this fraction for the E140H grating.  Each
data point represents the median value of the extracted gross spectrum
divided by the derived on-order background for a given spectral order
within a single observation, i.e., the median of
$b_\lambda/g_\lambda$.  In making this plot, we have extracted the
data in counts rather than flux.  This removes any differences between
the datasets caused by discrepant flux calibrations.  Average and
median values of the background to gross ratio are collected in Table
\ref{tab:e140hfrac} for several spectral orders.  In Figure
\ref{fig:e230hfrac} and Table \ref{tab:e230hfrac} we collect analogous 
values for data taken with the E230H grating.  All of the stars
included in these determinations are early type stars (O and B type,
as well as one white dwarf).  The ratio $b_\lambda/g_\lambda$ is
likely to depend on the shape of the underlying stellar continuum.
This is obvious even in the current data given the large dispersion
that appears near strong P Cygni stellar wind lines (e.g., \ion{Si}{4}
\wave{1400} and \ion{C}{4} \wave{1550} in Figure \ref{fig:e140hfrac}).

Figures \ref{fig:e140hfrac} and \ref{fig:e230hfrac} show a general
decrease in the prevalence of the background light at longer
wavelengths.  This is in qualitative agreement with expectation, since
most scattering phenomena show a decreasing intensity with increasing
wavelength.  For example, diffracted intensity decreases as
$\lambda^{-2}$, while Rayleigh scattering intensity decreases as
$\lambda^{-4}$.  Furthermore, the order spacing increases with
wavelength, making the fitting process more robust.  We find that the
background to gross fraction, $b_\lambda / g_\lambda$, decreases
approximately as $\lambda^{-3}$ for both gratings.  This scaling is
relationship is only approximate, and the E140H data are not well fit
by $b_\lambda / g_\lambda \propto \lambda^{-3}$ over the entire
wavelength range tested.  There is an excess in the $b_\lambda /
g_\lambda$ at short wavelengths compared with this scaling, and the
effects of strong stellar wind and interstellar absorption features
are visible in many other wavelength regions (see Figure
\ref{fig:e140hfrac}).  Furthermore, the ratio $b_\lambda /
g_\lambda$ for E230H data may rise beyond 3000 \AA, though our data
in that region are sparse.

The inter-order separation, $\Delta y_m$, in \stis\ high resolution
echelle-mode observations increases linearly with wavelength.  Hence
the background to gross fraction, $b_\lambda / g_\lambda$ can be
described as cubic polynomial of $1/\Delta y_m$.  Again, this scaling
is not perfect for E140H data.  Churchill \& Allen (1995) have shown a
similar correlation between the order separation and (in their case)
inter-order intensity, and they have used it to argue that the
contamination of adjacent orders $m\pm1$ (and beyond) was the most
important determiner of the on-order background spectrum in data from
the Hamilton Echelle Spectrograph.  However, we do not believe that
order overlap effects are causing the relationship seen in Figures
\ref{fig:e140hfrac} and \ref{fig:e230hfrac}.  The
lack of direct order overlap and the relatively small contribution
from the halos of adjacent orders (see Figure \ref{fig:psf}) in the
\stis\ data strongly suggests neither of these is driving the relationship 
seen in Figures \ref{fig:e140hfrac} and \ref{fig:e230hfrac}.
Potentially more likely is the effect of echelle scattering for small
order separations, but our discussion in \S \ref{subsec:limits}
suggests this contribution may not be enough to cause these
relationships either.  It is more likely that Figures
\ref{fig:e140hfrac} and \ref{fig:e230hfrac} are tracing the 
wavelength dependence of the processes causing the scattered and
background light.

The background to gross fractions shown in Figures \ref{fig:e140hfrac}
and \ref{fig:e230hfrac} are dependent on the size of the extraction
box used in determining the gross and background spectra.  The
cross-dispersion profile of the spectral trace is roughly Gaussian
(see Figure \ref{fig:xdisp}).  Our default extraction procedure adopts
a seven pixel extraction box (from $y_0-3$ to $y_0+3$, where $y_0$ is
the center of the spectral trace).  The outer-most pixels in the
extraction box have much higher background to gross values than the
inner-most pixels.  Thus, if only the central three pixels were used
in the spectral extraction (e.g., $y_0-1$ to $y_0+1$), the on-order
background to gross ratio would be significantly lower.  Experiments
using a three pixel extraction box for data near 1250 \AA\ suggests
the decrease in $b_\lambda / g_\lambda$ can be of order 30\%.

Given the relatively large contribution of the on-order background
light to the gross spectrum at short wavelengths, it may be better in
some cases to use smaller extraction boxes.  The size of the
extraction box can be chosen to minimize the impact of statistical
noise in the background on the extracted net spectrum.  If the
shortest wavelength region is vital to a \stis\ program, the
background fraction must be accounted for when calculating exposure
times.

\section{Comparisons with GHRS Data}
\label{sec:ghrs}

In order to assess the accuracy of our \stis\ background correction
scheme, we will compare \stis\ data extracted as described above with
data from the previous high-resolution UV spectrograph on-board \hst,
the Goddard High Resolution Spectrograph (GHRS).  This section
presents two such comparison datasets. First, we compare STIS E140H
observations of \hd 218915 with observations of the same star made
with the first-order G160M grating of the \ghrs.  The scattered light
properties of the intermediate resolution G160M grating are excellent
(see Cardelli \etal\ 1993), making data obtained with this grating a
good point of comparison for our background correction.  Second, we
compare archival STIS E140H observations of the nearby DA white dwarf
G191-B2B with \ghrs\ Ech-A high-resolution data for the same star.
The resolutions of the \stis\ and GHRS G191-B2B datasets are similar,
allowing a good comparison without the need to smooth the
\stis\ data.

\subsection{Comparison of \stis\ E140H and \ghrs\ G160M Observations of 
	\hd 218915}

The star \hd 218915 lies behind gas associated with the Perseus arm;
it was observed extensively with the \ghrs\ first-order G160M grating.
We have retrieved these data from the \hst\ archive and reduced them
as described by Howk, Savage, \& Fabian (1999).  These data, taken
through the small science aperture ($0\farcs25\times0\farcs25$) before
the installation of COSTAR, have a velocity resolution of $\sim18.6$
\kms\ (FWHM) at 1250 \AA.  Our comparison of the \stis\ and \ghrs\
data for this sightline will focus on the \ion{S}{2} lines
$\lambda1250$, 1253, and 1259 and \ion{Si}{2} $\lambda1260$; strong
\ion{C}{1} absorption is also seen in this spectral region (see Figure
\ref{fig:218915}).

The \stis\ E140H data, which have a resolution $\Delta v \sim 2.75$
\kms, must be smoothed for a direct comparison with the
intermediate-resolution \ghrs\ G160M data.  We have convolved the
higher-resolution \stis\ data, after combining several spectral
orders, with a Gaussian having a FWHM of 18.4 \kms\ and rebinned these
smoothed \stis\ data to approximately the same sampling as the \ghrs\
data.  We have also normalized the \stis\ data with a linear continuum
to bring the slope and continuum intensity levels of the \ghrs\ and
\stis\ data into agreement.  After the standard background
subtraction, we find no evidence for residual scattered light in the
\ghrs\ G160M observations, as judged by the saturated core of
\ion{Si}{2} $\lambda1260$.

\begin{figure}
\epsscale{0.85}
\plotone{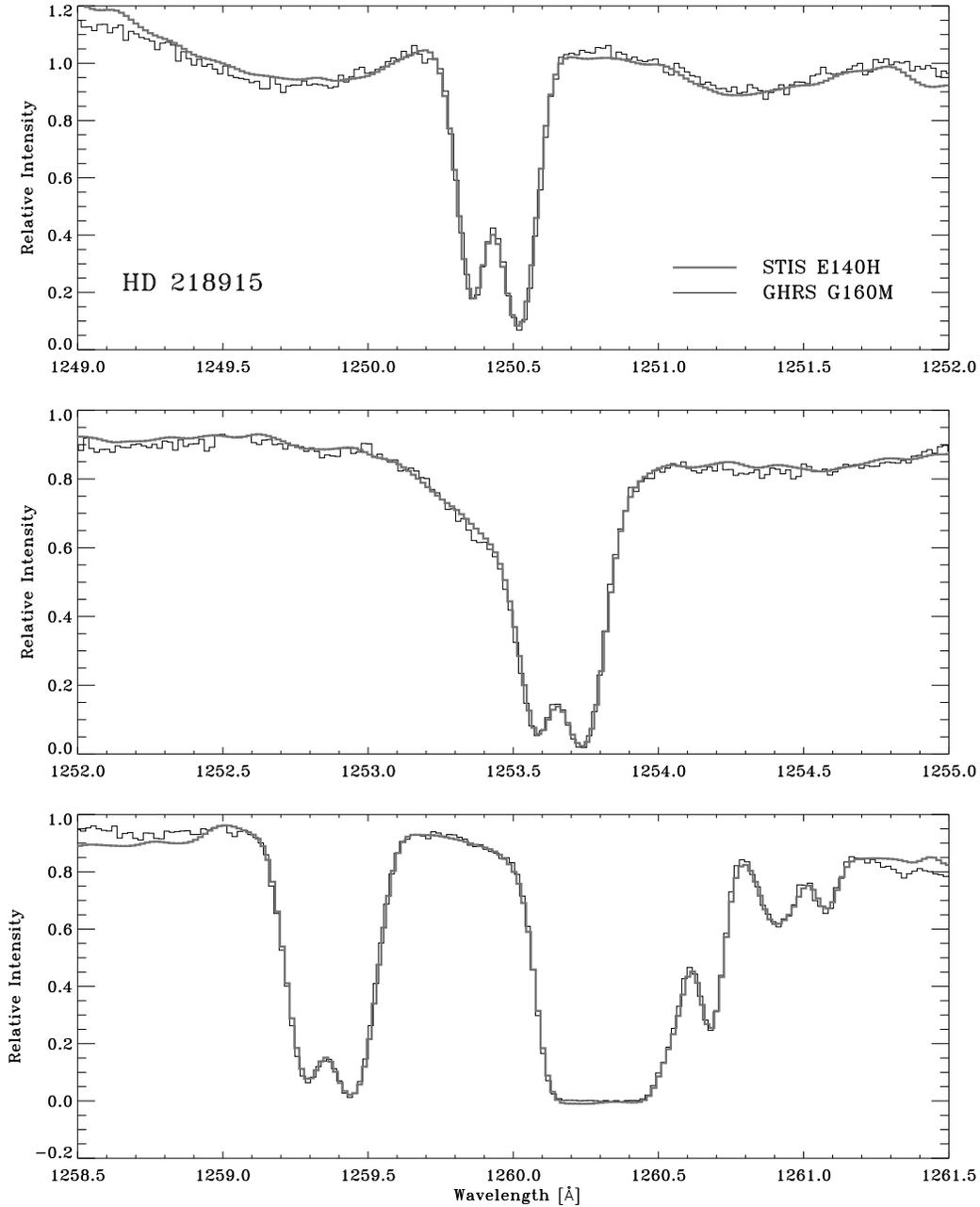}
\caption{A comparison of STIS E140H and GHRS G160M observations of the star
HD 218915 in three spectral regions containing \protect\ion{S}{2},
\protect\ion{Si}{2}, and \protect\ion{C}{1} interstellar absorption.  
The STIS data, displayed as a thick grey line, have been smoothed and
rebinned to the effective resolution of the GHRS data ($\sim18.6$
\protect\kms).  A rough linear continuum has been removed from 
the STIS data, and both datasets have been arbitrarily scaled by a
multiplicative constant to best match one another.  The agreement
between these two sets of observations is excellent at most
wavelengths.  A slight over-subtraction of the STIS data in the core
of \protect\ion{Si}{2} $\lambda1260$ is visible.  The agreement
between the two datasets is particularly encouraging given the good
scattered light properties of the GHRS G160M grating.  The smoothed
versions can be compared with the full-resolution STIS observations
shown in Figure
\ref{fig:218915}.  \label{fig:g160m}}
\end{figure}

Figure \ref{fig:g160m} shows the comparison between the \ghrs\ G160M
data and the smoothed \stis\ E140H data extracted with our background
correction scheme.  The top panel shows absorption due to \ion{S}{2}
$\lambda1250$, the middle panel shows \ion{S}{2} $\lambda1254$, and
the bottom panel includes absorption from \ion{S}{2} $\lambda1259$,
\ion{Si}{2} $\lambda1260$ (blended with \ion{Fe}{2}), and several
lines of \ion{C}{1}.  Unsmoothed data for the spectral region covered
in the bottom panel are shown in Figure \ref{fig:218915}.  We have
removed a slight velocity offset between the two datasets in producing
Figure \ref{fig:g160m}.

The agreement between the two datasets shown in Figure
\ref{fig:g160m} is generally excellent.  The depths and shapes of all 
three \ion{S}{2} lines are in good agreement between the \stis\ and
\ghrs\ datasets.  The \stis\ \ion{Si}{2} \wave{1260} profile exhibits
the slight oversubtraction of the saturated line core that we have
pointed out previously (on the low wavelength end).  However, the rest
of the \ion{Si}{2} profile and the adjoining \ion{C}{1} transitions
show remarkable agreement in the two datasets.  We find a similar
agreement between the longer wavelength transitions of \ion{Si}{2}
\wave{1526} and \ion{C}{4} \wave{1548}, though we do not show those
data.

Given the good scattered light properties of the first-order GHRS
G160M grating, the agreement between these two datasets suggests our
background subtraction algorithm is producing reliable estimates of
the on-order background spectrum, within the limits of this
comparison.  The heavy smoothing required to compare the \stis\ and
\ghrs\ datasets could potentially mask important differences in
the datasets.

\subsection{Comparison of \stis\ E140H and \ghrs\ Ech-A Observations of 
	G191-B2B}

The nearby ($d\sim69$ pc) DA white dwarf G191-B2B has been
well-observed by both the \stis\ and \ghrs\ in their high-resolution
modes.  The low column density of material along this sightline [$\log
N(\mbox{\ion{H}{1}}) \sim 18$ in atoms \column] has made it important
for studying the D/H abundance in the local ISM.  Results derived from
the \ghrs\ observations have been published by Vidal-Madjar \etal\
(1998).  The \stis\ data have been published by Sahu \etal\ (1999),
who used the STIS IDT background correction algorithm (Bowers \&
Lindler, in prep.) in their data reduction.

We retrieved the archival \ghrs\ Ech-A data for G191-B2B and reduced
them as described by Howk \etal\ (1999).  There are a large number of
\ghrs\ exposures covering the region of the spectrum containing
interstellar \lya.  We summed all those exposures taken at the same
grating carrousel position within an individual observation (visit).
We then merged the resulting vectors using the standard wavelength
scales, at the same time solving for and removing the fixed-pattern
noise spectrum.  The standard \ghrs\ background subtraction
over-estimates the background level near the saturated interstellar
\lya\ profile.  We adopted a value for the ``$d$-coefficient'' of
$d=0.002$ to bring this saturate profile to the appropriate zero-level
(see Cardelli \etal\ 1990, 1993).\footnote{That this $d$-coefficient
is lower than that usually predicted near \lya\ is not unexpected
given the much higher continuum levels of G191-B2B in this region of
the spectrum compared with stars behind larger interstellar column
densities.} All of the other data presented here have been reduced
using the standard $d$ values of Cardelli \etal\ (1993).  The G191-B2B
\ghrs\ data presented here were all taken through the small science
aperture.  The \ion{Si}{3} profile lies at the edge of the Digicon
detector array in the \ghrs\ observations of the 1206 \AA\ region.  We
have used only the three (of four) FP-SPLIT positions where the
\ion{Si}{3} absorption was shifted away from the edge of the detector array
in deriving the profile seen in the bottom panel of Figure
\ref{fig:g191}.

\begin{figure}
\epsscale{0.99}
\plotone{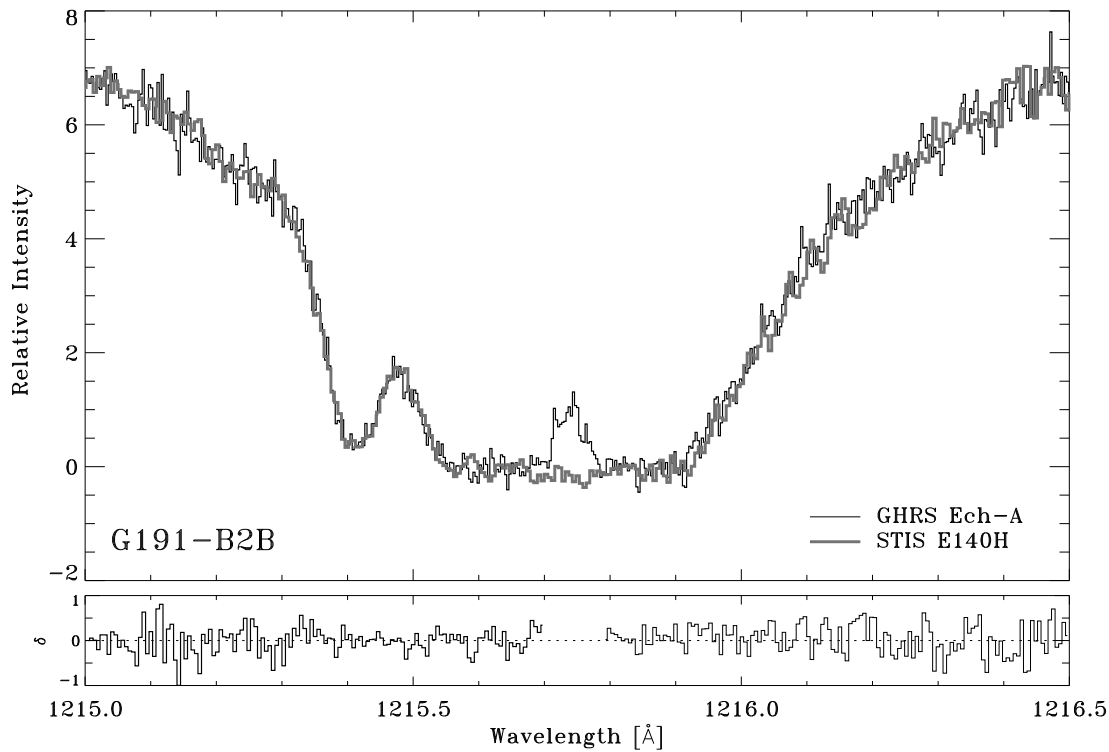}
\caption{A comparison of the interstellar \protect\ion{H}{1} and 
\protect\ion{D}{1}  Lyman-$\alpha$ observations towards the DA white 
dwarf G191-B2B using STIS and GHRS.  This figure can be directly
compared with Figure 3 of Sahu \etal\ (1999).  The post-COSTAR GHRS
data have slightly lower resolution but higher sampling.  Our
extractions of the STIS data are shown by the thick grey line.  We
have coadded two orders (346 and 347) in our STIS extraction.  The
bottom panel shows the difference between the GHRS data (after
rebinning to the same sampling as the STIS data) and the STIS data.
The GHRS data include airglow emission in the center of the Ly$\alpha$
profile that is not present in the STIS data; the residuals from the
region containing this emission are not shown.  The residuals between
the two datasets are completely consistent with the formal STIS errors
(71\% of the residual points fall within $1\sigma$, while 95\% of the
residual points fall within $2\sigma$).  There is no compelling reason
to believe that the \stis\ and \ghrs\ data along this sightline are in
disagreement as suggested by Sahu \etal\ (1999).  Our \ghrs\
\lya\ profile is in good agreement with that of Vidal-Madjar \etal\
(1998).
\label{fig:HI}}
\end{figure}

\begin{figure}
\epsscale{0.7}
\plotone{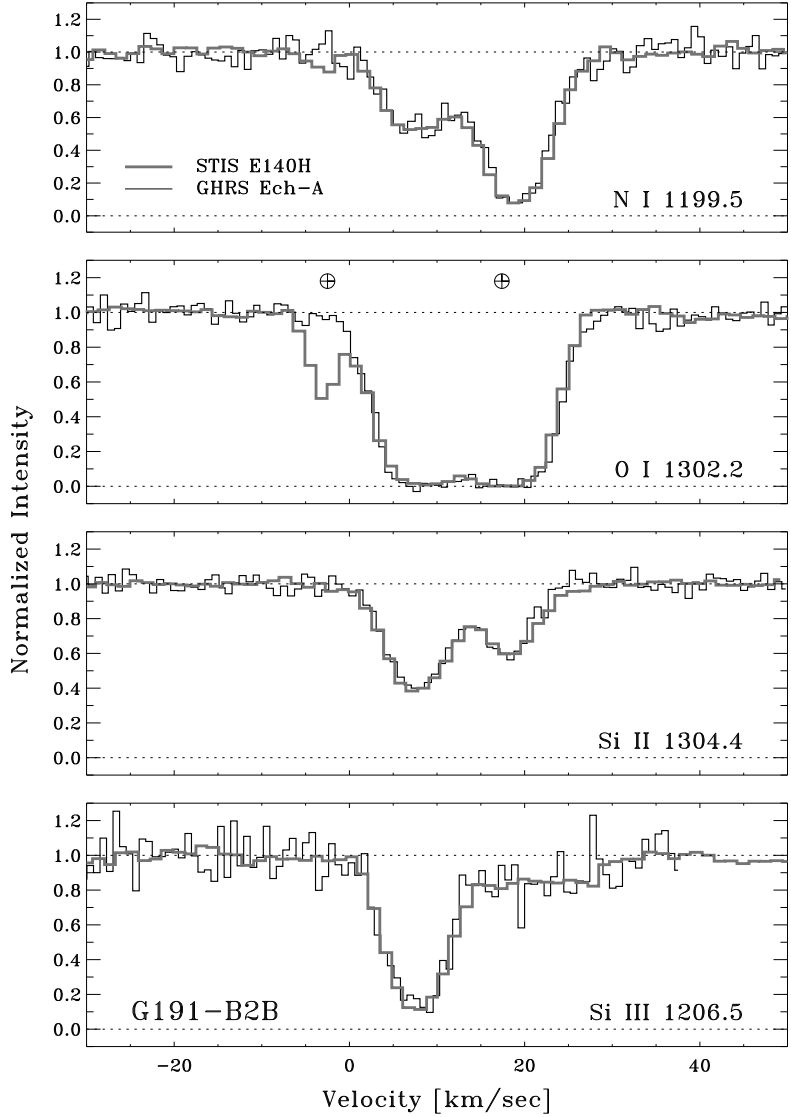}
\caption{A comparison of archival STIS E140H and GHRS Ech-A
observations of interstellar absorption lines towards the DA white
dwarf G191-B2B.  We show normalized absorption line profiles as a
function of velocity for the ions \protect\ion{N}{1},
\protect\ion{O}{1}, \protect\ion{Si}{2}, and \protect\ion{Si}{3}.  The
STIS data are shown as a thick grey line.  We have marked the
positions of telluric absorption lines in the \protect\ion{O}{1}
profile ($v\sim-2.5$ \protect\kms\ for the STIS observations and
$v\sim17.4$ \protect\kms\ for the GHRS observations).  The GHRS
\ion{Si}{3} observations are at the end of the detector diode array, 
making the continuum placement uncertain.  In general we find
excellent agreement between these two datasets.
\label{fig:g191}}
\end{figure}

Figure \ref{fig:HI} shows a comparison of our extraction of the \stis\
E140H observations of the wavelength region surrounding \lya\ with the
\ghrs\ Ech-A observations of this line.  We have
co-added two overlapping orders to produce the \stis\ profiles (orders
346 and 347).  The \ghrs\ data have been scaled by a multiplicative
constant to match the \stis\ data.  The velocity resolution of the
\ghrs\ Ech-A data ($\sim3.5$ \kms) is somewhat worse than the
resolution of the \stis\ E140H data; however, the \ghrs\ data have
finer sampling ($\sim0.88$ \kms\ vs. $\sim1.25$ \kms\ for the \stis\
E140H data).  Figure 3 of Sahu \etal\ (1999) compares the STIS \lya\
profile as extracted with the IDT reduction routines with their
reduction of the \ghrs\ Ech-A data (though they have applied a
three-pixel smoothing kernel to their data).

Again, the agreement between the \ghrs\ and \stis\ profiles shown here
is excellent.  The curves are virtually indistinguishable, save for
the geocoronal \lya\ emission in the center of the \ghrs\ profile.
The center of the \stis\ \lya\ profile shows a slight
over-subtraction, consistent with the discussion in \S
\ref{sec:examples}.  The \ion{D}{1} \lya\ profiles are in 
excellent agreement.  There is no compelling reason to believe that
the \stis\ and \ghrs\ data along this sightline are in disagreement as
suggested by Sahu \etal\ (1999).  Our \ghrs\ \lya\ profile is in good
agreement with that of Vidal-Madjar \etal\ (1998).  Our profile is in
agreement with that derived by the IDT reduction software {\em to
within the statistical and background uncertainties.}


Figure \ref{fig:g191} shows a comparison of the normalized profiles
(as a function of velocity) for several interstellar absorption lines
extracted from the \stis\ and \ghrs\ observations of G191-B2B.  We
have normalized the data using low-order polynomial fits (see Howk
\etal\ 1999).  The \ghrs\ \ion{Si}{3} profile is taken from the 
very end of the order, and the continuum is uncertain for this
profile.  The \ion{O}{1} (and possibly also \ion{N}{1}) absorption
includes a telluric contribution.  The expected velocities of the
telluric absorption in the \stis\ and \ghrs\ data are marked.  The
atmospheric \ion{O}{1} is cleanly separated from the interstellar
absorption in the \stis\ data.

Again we see the agreement between \ghrs\ and \stis\ observations of
the same absorption lines is excellent.  There is no compelling
reason, given the comparisons shown in Figures \ref{fig:HI} and
\ref{fig:g191}, to believe the earlier \ghrs\ data provide different
results than the \stis\ data for this sightline as suggested by Sahu
\etal\ (1999).

\section{Summary and Recommendations}
\label{sec:summary}

We have presented a simple approach to estimating the on-order
background spectrum in the echelle modes of \stis.  Our algorithm fits
the cross-dispersion profile of the inter-order light and uses this fit
to estimate the on-order background.  The resulting
background-corrected net spectra show strong interstellar lines with
zero residual flux, as expected.  The most important aspects of our
algorithm, and of the \stis\ echelle background in general, are as
follows.

\begin{enumerate}
\item STIS echelle data contain significant amounts of scattered
light.  The amount of scattered light present depends upon wavelength.
The ratio of the on-order background to gross spectra, $b_\lambda /
g_\lambda$, varies as roughly $\lambda^{-3}$ and ranges from $\sim
0.1$ at long wavelengths to $\sim 0.5$ at short wavelengths.

\item The effectiveness of the background-correction algorithm
presented here depends upon wavelength and the signal-to-noise ratio
of the spectrum being analyzed.  Short wavelength \stis\ echelle data,
$\lambda \sim 1300$ \AA, often show some residual artifacts in the
cores of saturated interstellar absorption lines, though the
difficulties are most important for wavelengths shortward of \lya.
This is in part due to the close spacing of the spectral orders in
\stis\ echelle mode observations at short wavelengths.  Low
signal-to-noise ratio data cause greater difficulties in the
background fitting process.  The over-subtraction of portions of short
wavelength saturated lines is typically $\la 1-2\%$ of the local
continuum.


\item Scattered light and the uncertainties involved in removing it
can affect spectral line analyses and must be taken into account when
analyzing lines with substantial optical depths.  The potential
systematic uncertainties in the optical depths for lines with
(non-zero) residual fluxes $\la 10\%$ of the local continuum can be
$\ga10\%$.

\item Comparisons of high-resolution \stis\ data with \ghrs\ high- and
intermediate-resolution data show no evidence for a significant
difference between \stis\ data reduced with our background subtraction
technique and \ghrs\ data using standard background-subtraction
techniques for that instrument (Cardelli \etal\ 1990, 1993).

\end{enumerate}

We believe our data extraction routine provides reliable net spectra
for the high-resolution modes of \stis, within the framework of the
caveats and problems outlined in this work.  However, we believe that
much work still needs to be done to fully understand the
\stis\ background.  Given the possible importance of a reduction
scheme such as this, even though it is admittedly flawed in some
respects, we will make our IDL extraction routines available to the
general astronomical community.  Please contact the authors for more
information.

\acknowledgements

We appreciate useful comments from C. Churchill, B. Savage, and
E. Jenkins.  We thank A. Vidal-Madjar for allowing us to compare
\ghrs\ \lya\ profile for G191-B2B with his own.  We also appreciate
J. Valenti for allowing us to test our procedure on his \stis\ data
for $\alpha$ Cen.  We acknowledge support from NASA Long Term Space
Astrophysics grant NAG5-3485 and grant GO-0720.01-96A from the Space
Telescope Science Institute, which is operated by the Association of
Universities for Research in Astronomy, Inc., under NASA contract
NAS5-26555.



\begin{deluxetable}{ccccc}
\tablenum{1}
\tablecolumns{5}
\tablewidth{0pc}
\tablecaption{Background Fractions for the STIS E140H\label{tab:e140hfrac}}
\tablehead{
\colhead{} & \colhead{} & \colhead{} &
\multicolumn{2}{c}{Background/Gross} \\
\cline{4-5}
\colhead{Order}  & 
\colhead{$\lambda_c$\tablenotemark{a}}  & 
\colhead{No.\tablenotemark{b}} & 
\colhead{Mean}  & 
\colhead{Median}
}
\startdata
  360  &  1169  &  9  &  $0.50 \pm 0.01$  & 0.49  \nl 
  355  &  1185  &  9  &  $0.43 \pm 0.01$  & 0.43  \nl 
  350  &  1202  &  9  &  $0.39 \pm 0.01$  & 0.39  \nl 
  340  &  1238  &  9  &  $0.33 \pm 0.02$  & 0.33  \nl 
  335  &  1256  &  9  &  $0.29 \pm 0.01$  & 0.29  \nl 
  330  &  1275  &  9  &  $0.28 \pm 0.01$  & 0.28  \nl 
  325  &  1295  &  9  &  $0.26 \pm 0.01$  & 0.26  \nl 
  320  &  1315  &  9  &  $0.23 \pm 0.01$  & 0.23  \nl 
  315  &  1336  & 10  &  $0.22 \pm 0.01$  & 0.22  \nl 
  310  &  1358  &  9  &  $0.21 \pm 0.01$  & 0.21  \nl 
  305  &  1380  &  5  &  $0.19 \pm 0.02$  & 0.18  \nl 
  300  &  1403  &  5  &  $0.19 \pm 0.02$  & 0.19  \nl 
  295  &  1427  &  5  &  $0.18 \pm 0.01$  & 0.18  \nl 
  290  &  1451  &  5  &  $0.16 \pm 0.01$  & 0.16  \nl 
  285  &  1477  &  5  &  $0.16 \pm 0.01$  & 0.16  \nl 
  280  &  1503  &  6  &  $0.15 \pm 0.01$  & 0.15  \nl 
  275  &  1531  &  5  &  $0.16 \pm 0.02$  & 0.16  \nl 
\enddata		        		       
\tablenotetext{a}{The central wavelength of the order.}
\tablenotetext{b}{The number of observations used in calculating
	the data for each order.}
\end{deluxetable}				       

\pagebreak

\begin{deluxetable}{ccccc}
\tablenum{2}
\tablecolumns{5}
\tablewidth{0pc}
\tablecaption{Background Fractions for the STIS E230H\label{tab:e230hfrac}}
\tablehead{
\colhead{} & \colhead{} & \colhead{} &
\multicolumn{2}{c}{Background/Gross} \\
\cline{4-5}
\colhead{Order}  & 
\colhead{$\lambda_c$\tablenotemark{a}}  & 
\colhead{No.\tablenotemark{b}} & 
\colhead{Mean}  & 
\colhead{Median}
}
\startdata
  470  &   1642  &   3   &   $0.40 \pm 0.17$   &   0.33  \nl
  460  &   1678  &   3   &   $0.27 \pm 0.03$   &   0.27  \nl
  450  &   1716  &   3   &   $0.28 \pm 0.03$   &   0.28  \nl
  440  &   1755  &   3   &   $0.25 \pm 0.05$   &   0.26  \nl
  430  &   1796  &   3   &   $0.23 \pm 0.05$   &   0.21  \nl
  420  &   1839  &   3   &   $0.19 \pm 0.02$   &   0.19  \nl
  410  &   1884  &   5   &   $0.19 \pm 0.02$   &   0.18  \nl
  400  &   1930  &   2   &   $0.21 \pm 0.04$   &   0.24  \nl
  390  &   1980  &   2   &   $0.18 \pm 0.04$   &   0.21  \nl
  380  &   2032  &   2   &   $0.15 \pm 0.01$   &   0.16  \nl
  370  &   2087  &   2   &   $0.14 \pm 0.01$   &   0.15  \nl
  360  &   2144  &   5   &   $0.13 \pm 0.02$   &   0.14  \nl
  350  &   2206  &   3   &   $0.12 \pm 0.02$   &   0.12  \nl
  340  &   2271  &   3   &   $0.11 \pm 0.02$   &   0.11  \nl
  330  &   2340  &   3   &   $0.12 \pm 0.03$   &   0.11  \nl
  320  &   2412  &   3   &   $0.08 \pm 0.01$   &   0.08  \nl
  310  &   2491  &   3   &   $0.09 \pm 0.01$   &   0.09  \nl
  300  &   2574  &   3   &   $0.09 \pm 0.02$   &   0.08  \nl
  290  &   2662  &   5   &   $0.08 \pm 0.01$   &   0.08  \nl
  280  &   2758  &   6   &   $0.01 \pm 0.05$   &   0.07  \nl
  270  &   2860  &   6   &   $0.08 \pm 0.02$   &   0.09  \nl
  260  &   2969  &   3   &   $0.08 \pm 0.02$   &   0.08  \nl
\enddata
\tablenotetext{a}{The central wavelength of the order.}
\tablenotetext{b}{The number of observations used in calculating 
the data for each order.}
\end{deluxetable}

\clearpage


\clearpage 

\clearpage

\clearpage

\clearpage

\clearpage

\clearpage

\end{document}